\documentclass[prd,nofootinbib,preprint,superscriptaddress]{revtex4-1}
\usepackage{amsmath}
\usepackage{graphics}
\usepackage{epstopdf}
\usepackage{graphicx}
\usepackage{subfigure}

\makeatletter
\renewcommand{\fnum@table}{\textbf{\tablename~\thetable}}
\renewcommand{\fnum@figure}{\textbf{\figurename~\thefigure}}
\makeatother

\newcommand {\be}{\begin{equation}}
\newcommand {\ee}{\end{equation}}
\newcommand {\ba}{\begin{eqnarray}}
\newcommand {\ea}{\end{eqnarray}}

\begin{document}


\vspace*{10mm}

\title{Sensitivities of future reactor and long-baseline neutrino experiments to NSI \vspace*{1.cm} }

\author{\bf Pouya Bakhti}
\affiliation{Institute for
	research in fundamental sciences (IPM), PO Box 19395-5531, Tehran,
	Iran}
\author{\bf Meshkat Rajaee}
\affiliation{Institute for
	research in fundamental sciences (IPM), PO Box 19395-5531, Tehran,
	Iran}

\begin{abstract}
  \vspace*{.5cm}

We investigate the potential of the next generation long-baseline neutrino experiments DUNE and T2HK as well as the upcoming reactor experiment JUNO to constrain Non-Standard Interaction (NSI) parameters. JUNO is going to provide the most precise measurements of solar neutrino oscillation parameters as well as determining the neutrino mass ordering.
We study how the results of JUNO combined with those of long-baseline neutrino
experiments such as DUNE and T2HK can help to determine oscillation parameters and to constrain NSI parameters. We present excluded regions in NSI parameter space, $\epsilon_{\alpha \beta}$ assuming Standard Model (SM) as the null hypothesis. We further explore the
correlations between the NSI parameters and CP-violation phase.

\end{abstract}

\maketitle

\section{Introduction}

Neutrino oscillations can successfully describe neutrino
flavor transitions. In 1978, the matter potential for neutrino propagation
through matter
was proposed \cite{Wolf}. Wolfenstein introduced non-standard interactions besides the neutrino mass matrix
as another approach to probe new physics beyond the standard model \cite{Wolf}.

The NSI describes a large class of new physics models where the neutrino interactions with ordinary matter are parameterized at low energy in terms of effective flavor-dependent couplings $\epsilon_{\alpha \beta}$.
Several models of physics beyond the SM predict NSI. For instance, NSI
arise naturally in many neutrino mass models trying to explain the smallness of
neutrino mass \cite{ns1,ns2,ns3,ns4,ns5,ns6,ns7} and the large neutrino mixing angles \cite{mixing1,mixing2,mixing3}. Thus, it is crucial to understand how the presence of NSI can affect the standard neutrino oscillation in matter.
The presence of NSI couplings can affect neutrino production, detection and propagation.
Such a new interaction leads to a rich phenomenology and has been extensively studied in the literature. NSI were proposed as a solution to
the solar neutrino problem \cite{Sol}. Also, see references \cite{sol1,sol2,sol3,sol4,sol5,Bakhti:2020hbz} for studies of their impact on solar neutrinos oscillations.
NSI impacts on accelerator neutrinos and atmospheric neutrinos
have been explored in \cite{ac1,ac2,ac3,ac4,ac5,ac6,ac7,ac8,atm1,atm2,atm3,atm4,atm5,atm6,atm7,atm8,atm9,atm10,atm11,atm12,atm13,Bakhti:2016prn}

NSI can be formulized as a $d=6$ four-fermion effective operators involving neutrino fields given in two main categories:
Operators affecting charged-current neutrino interactions $(\bar l_\alpha \gamma_\mu P_L \nu_\beta) (\bar q \gamma^\mu P q')$, where $l$ stands for a charged lepton, $P$ stands for one of the chirality projectors $P_{R,L} \equiv\frac{1}{2}(1 \pm \gamma_5)$, $\alpha$ and $\beta$ are lepton flavor indices, and $q$ and $q'$ represent quarks,
or operators affecting neutral-current neutrino interactions which are given in the form $(\bar\nu_\alpha \gamma_\mu P_L \nu_\beta) (\bar f \gamma^\mu P f)$, where $f$ represents SM fermion.
Charged-current neutrino interactions NSI affect neutrino production and detection processes.
See Refs.~ \cite{Bakhti:2016gic,nsi1,nsi2,nsi3,nsi4,nsi5,nsi6,nsi7,nsi8,nsi9,nsi10}
for the studies on the potential of neutrino oscillation experiments to study NSI affecting neutrino production and detection.
On the other hand, neutral-current NSI affects neutrino propagation.
Since the present bounds on the CC NSI parameters are strong while the constraints on NC NSI are not very stringent, in this paper we focus on the neutral-current neutrino nonstandard interactions.

Interestingly, matter effects are negligible for the reactor experiments due to their low energy and relatively short-baseline. This also holds for NSI effects during propagation of neutrinos and therefore, we can only assume the NSI effects only at sources and detectors. Using this characteristic of reactor experiments, we focus on investigating how
well the future long-baseline neutrino experiments can constrain non-standard neutrino interaction.
The exact value of the mixing angle
$\theta_{13}$
has now been determined to good accuracy by reactor experiments such as Daya Bay \cite{dayabay}
and RENO \cite{reno} with the previous hint from combination of KamLAND and solar neutrino observatories data \cite{Gando:2010aa}. These experiments provided
evidence for a relatively large angle $\theta_{13}$, with 5.2$\sigma$ and
4.9$\sigma$ results respectively and their results show that $\theta_{13}$ is very close to 8.4$^\circ$.
Besides, the T2K experiment
has also confirmed the non-zero and moderately
large value of $\theta_{13}$ in the standard three-flavor oscillation scenario \cite{t2k}. The future reactor experiment Jiangmen Underground Neutrino Observatory (JUNO) is going to determine the mass ordering and is going to measure $\Delta m ^2 _{21} $ and $\theta _{12} $ to the percent level \cite{Jun}.
JUNO is a multipurpose neutrino experiment designed to detect solar neutrinos and geo-neutrinos in addition to determining neutrino mass hierarchy and precisely measurement of oscillation parameters.

The long-baseline Deep Underground Neutrino Experiment (DUNE) with the main goal of searching for leptonic CP-violation phase and testing the three massive neutrinos paradigm is going to provide precision measurements of oscillation probabilities which can be used to probe neutral-current interactions (NSI).
The effects of NSI and how they modify neutrino propagation in the DUNE is studied in \cite{degouvea}.
The capability of the DUNE Near Detector (ND) to constrain Non-Standard
Interaction parameters (NSI) describing the production of neutrinos is studied in
\cite{Giarnetti:2020bmf,Bakhti:2016gic}.
The Tokai to Hyper-Kamiokande (T2HK) experiment uses an upgraded J-PARC beam with a detector
located 295 km away from the source with a spectrum peaked at 0.6 GeV \cite{t2h}.
The effect of the presence of NSI in DUNE, T2HK and T2HKK has been studied in \cite{liao}.
References \cite{Adhikari:2012vc, Girardi:2014kca} present the Daya Bay and T2K results on and non-standard neutrino interactions.

In this paper, we study how combining data of future long-baseline experiments DUNE and T2HK with the precise measurements of future reactor experiment JUNO is going to determine the standard oscillation parameters by higher accuracy. Moreover, we present the constraints on NSI parameter space using these future neutrino experiments. In more details, we study how well the precise measurements of oscillation probabilities at DUNE, T2HK and JUNO can be used to probe
the existence of non-standard neutrino neutral-current-like interactions and discuss how using
combined data of the upcoming DUNE, T2HK and JUNO experiments can constrain NSI assuming it is consistent with the standard paradigm. We will investigate the
correlations between the NSI parameters and CP-violation phase using our simulations for these future neutrino experiments.

The paper is organized as follows. In Sec. II, we discuss the NSI Lagrangian and its effect
on neutrino oscillation. In Sec. III, we discuss the details of our simulation and the experiments. In Sec. IV, we present our results. Our conclusions will be presented in section V.

\section{Formalism}

In this section, we briefly discuss how the presence of NSI modifies
the effective Hamiltonian for neutrino propagation
in the matter. The total Hamiltonian in the flavor basis is given by \cite{ponte}
\begin{equation}
H \;= H_{vac} + H_{matt} = \; \dfrac{1}{2E}
\left(
U
\begin{bmatrix}
m_1^2 & 0 & 0 \\
0 & m_2^2 & 0 \\
0 & 0 & m_3^2
\end{bmatrix}
U^\dagger
+
a
\begin{bmatrix}
1 + \varepsilon_{ee} & \varepsilon_{e\mu} & \varepsilon_{e\tau} \\
\varepsilon_{e\mu}^* & \varepsilon_{\mu\mu} & \varepsilon_{\mu\tau} \\
\varepsilon_{e\tau}^* & \varepsilon_{\mu\tau}^* & \varepsilon_{\tau\tau}
\end{bmatrix}
\right) \label{matrix}
\end{equation}
where $E$ is the neutrino
energy and $a$ is matter-effect parameter and is given by
\begin{equation}
a
\;=\; 2\sqrt{2}G_F N_e E
\;=\; 7.6324\times 10^{-5}(\mathrm{eV}^2)
\left(\dfrac{\rho}{\mathrm{g/cm^3}}\right)
\left(\dfrac{E}{\mathrm{GeV}}\right)
\;,
\label{matter-term}
\end{equation}

and the ``$+1$'' term in Eq.~\ref{matrix} corresponds to the standard
contribution, and
\begin{equation}
\label{eq:epx-nsi}
\epsilon_{\alpha\beta} = \sum_{f=e,u,d}
\frac{N_f(x)}{N_e(x)} \epsilon_{\alpha\beta}^f
= \big[ 4 + 3 Y_n(x) \big] \epsilon_{\alpha\beta}^f \,,
\end{equation}
with
\begin{equation}
Y_n(x) \equiv \frac{N_n(x)}{N_e(x)}
\end{equation}
describes the non-standard part and is written in terms of the
effective couplings of protons ($p$) and neutrons ($n$). We have assumed $N_e(x) = N_p(x) = N_n(x)$ so $Y_n(x) = 1 $.

$U$ is the Pontecorvo-Maki-Nakagawa-Sakata (PMNS) matrix and is parameterized as
\begin{align}
U = \begin{pmatrix}
c_{12} c_{13} & s_{12} c_{13} & s_{13} e^{-i\delta_{\rm CP}} \\
-s_{12} c_{23} - c_{12} s_{13} s_{23} e^{i\delta_{\rm CP}} &
c_{12} c_{23} - s_{12} s_{13} s_{23} e^{i\delta_{\rm CP}} & c_{13} s_{23} \\
s_{12} s_{23} - c_{12} s_{13} c_{23} e^{i\delta_{\rm CP}} &
-c_{12} s_{23} - s_{12} s_{13} c_{23} e^{i\delta_{\rm CP}} & c_{13} c_{23}
\end{pmatrix}.
\label{eq:UPMNS}
\end{align}
where $s_{ij}$ and $c_{ij}$ denote the sine and cosine of the mixing angle
$\theta_{ij}$, and $\delta_{\rm CP}$ is the (Dirac) CP phase \cite{ponte,maki},
The strongest bounds on NSI parameters
in propagation come from the global fit to neutrino oscillation data in Ref.~\cite{Esteban:2019lfo}. In presence of NSI in propagation, global
analyses of neutrino oscillation data are compatible with two solutions: the LMA solution and e LMA-dark solution. We focus on LMA solution since the LMA dark solution is excluded by the COHERENT experiment \cite{Coloma:2017ncl} at 3$\sigma$ C.L..

 The oscillation probability in the presence of NSI is given by \cite{kopp}
\begin{align}
  P_{\nu _\alpha \rightarrow \nu _\beta}
    &= |< \nu_\beta \mid  e^{-i H L}  \mid \nu_\alpha >|^2  
  \label{eq:P-ansatz}
\end{align}

Since reactor experiments are sensitive to 
$P(\bar{\nu}_e \rightarrow \bar{\nu}_e)$  and long-baseline experiments are sensitive to 
$P( \overset{\scriptscriptstyle(-)}{ \nu } _ {\mu}  ~  \rightarrow ~\overset{\scriptscriptstyle(-)}{\nu } _e )$ 
and $P (   \overset{\scriptscriptstyle(-)}{ \nu } _ \mu ~ \rightarrow ~ \overset{\scriptscriptstyle(-)}{ \nu } _ \mu )$,
 in order to study the impact of non-standard interactions on 
these experiments, we only  need to focus on these   oscillation probabilities.
Since the oscillation probability $P(\nu_\mu \rightarrow \nu_e)$ 
does not have strong dependence on $\mid \epsilon_{\mu \tau} \mid $ and  $\mid  \epsilon_{\mu \mu} - \epsilon_{\tau \tau} \mid $ and there are already strong bounds on $\mid \epsilon_{\mu \tau} \mid $ and on $\mid  \epsilon_{\mu \mu} - \epsilon_{\tau \tau} \mid $ \cite{Esteban:2019lfo}, we take  $  \epsilon_{e \mu}  $, $  \epsilon_{e \tau}   $ and   $   \epsilon_{ee} - \epsilon_{\mu \mu}   $ non-zero. 
 To perform our analysis, we set true values of standard oscillation parameters from nu-fit \cite{Esteban:2020cvm} and the uncertainties from \cite{Esteban:2019lfo}. 
Also, we set the true values of  $  \epsilon_{e \mu}  $, $  \epsilon_{e \tau}   $ and   $   \epsilon_{ee} - \epsilon_{\mu \mu}   $ to zero.

\section{ DETAILS OF the ANALYSIS and RESULTS}

In this section, we discuss the approach of our
analysis in obtaining the results presented in this paper. For the statistical inferences, we simulate 10 years of data collection for all of the experiments. For DUNE and T2HK, we have considered data taking for 5 years in each mode.

We have considered that DUNE consists of a $40$ kiloton liquid argon detector and utilizes a $1.2$ MW proton beam to produce neutrino and antineutrino beams from in-flight pion decay, originating $1300$ km upstream at Fermilab. The neutrino energy ranges between $0.5$ and $20$ GeV and the flux peak occurs around $3.0$ GeV. All the details of the DUNE experiment are taken from \cite{Bakhti:2016gic, Acciarri:2015uup}

For the T2HK experiment, we have considered the mass of $225$ kt for the water Cherenkov detector \cite{Hyper-Kamiokande:2016dsw}.
T2HK experiment uses an upgraded 30 GeV J-PARC beam with a power
of 1.3 MW and its detector is
located 295 km away from the source. All the details of the T2HK experiment can be found in \cite{t2hk, Hyper-Kamiokande:2016dsw, Bakhti}. Medium-baseline reactor experiment JUNO consists of nuclear plant reactor complex at Yangjiang and Taishan with the power of 36 GW and a 20 kt scintillator detector with a baseline of 52 km. We have considered all the details of JUNO the same as given in \cite{Bakhti:2013ora, Bakhti:2014pva} and we have assumed 10 years of data taking for JUNO.

For our analysis, we have considered the value of $\theta_{13}$ as measured by the Daya Bay experiment.
As mentioned in the previous section, we set true values of standard oscillation parameters from nu-fit \cite{Esteban:2020cvm} and the uncertainties from \cite{Esteban:2019lfo} while setting the true values of $ \epsilon_{e \mu} $, $ \epsilon_{e \tau} $ and $ \epsilon_{ee} - \epsilon_{\mu \mu} $ equal to zero. We marginalize over phases of $ \phi_{e \mu} $, $ \phi_{e \tau} $ and other systematic using the so-called pull method. In the following, we will explore the future sensitivity of the combination of DUNE, T2HK and JUNO to determine standard neutrino oscillation parameters in the presence of NSI and try to constrain NSI parameters.
For the statistical inferences we have considered Asimov data set approximation,
and we have assumed the standard model as the true model. We have used chi-squared method with the assumption of Gaussian distribution of errors.
In our simulation, we used GLoBES software \cite{Huber:2004ka, Huber:2007ji} and include the NSI from Ref.~\cite{kopp}.
Let us discuss the main results of the effects of NC NSIs on the determination of the standard oscillation parameters considering JUNO, DUNE and T2HK experiments.

\begin{figure}[h]
\hspace{0cm}
\includegraphics[width=0.49\textwidth, height=0.37\textwidth]{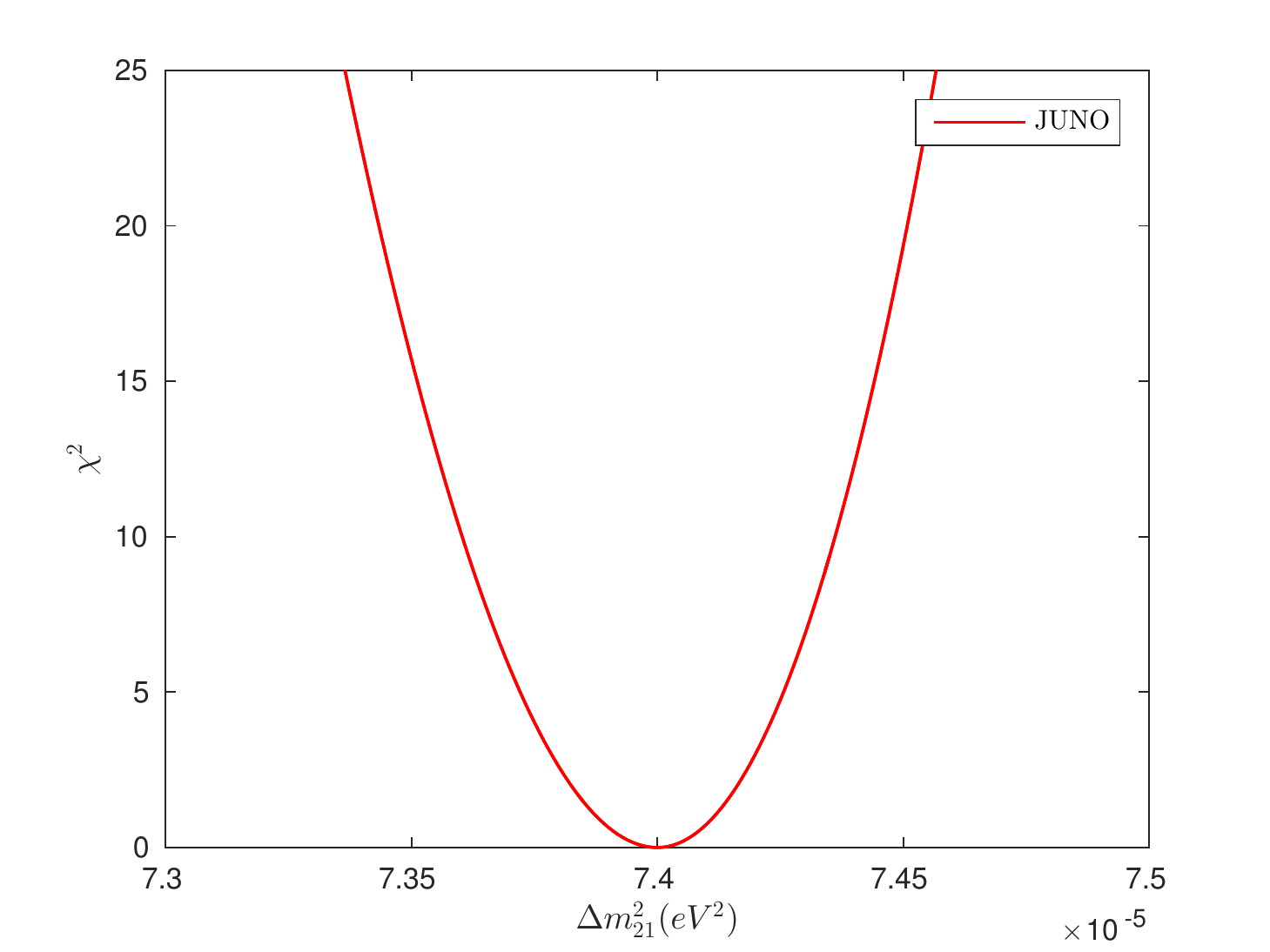}
\includegraphics[width=0.49\textwidth, height=0.37\textwidth]{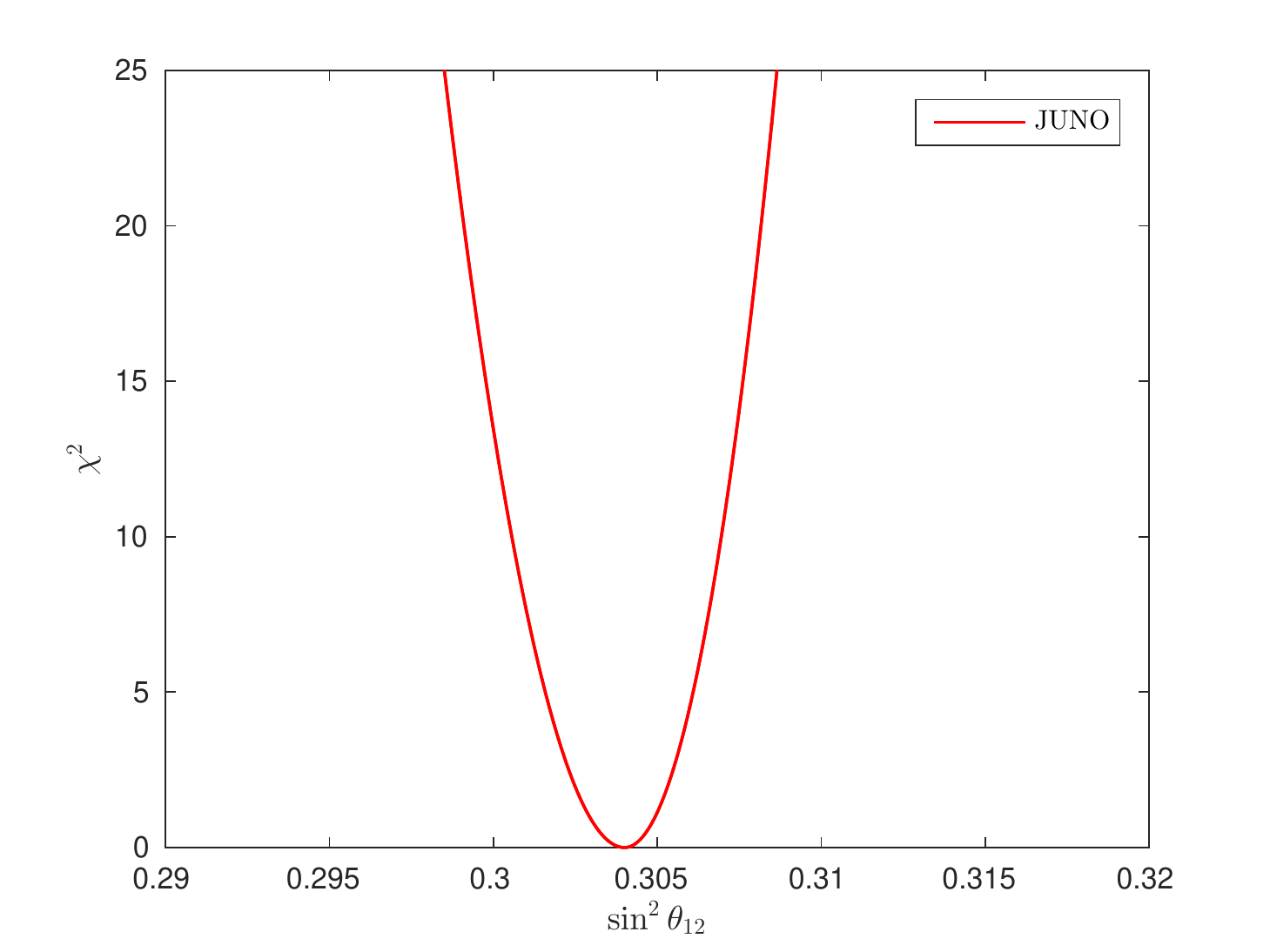}
\caption[...]{
 Chi-squared vs. $\Delta m^2 _{21}$ and $\sin^2\theta_{12}$ assuming ten years of data taking for the JUNO experiment. We have set the true values from nu-fit \cite{Esteban:2020cvm}.
\label{dm21}
}
\end{figure}

Fig.~\ref{dm21} shows the sensitivity of JUNO to the measurement of solar neutrino oscillation parameters $\Delta m^2 _{21}$ and $\theta_{12}$. As it was expected NSI does not affect JUNO measurements on the oscillation parameters as studied in \cite{Bakhti:2020hbz}. Thus, in the presence of NSI, JUNO can measure these oscillation parameters by high precision.

\begin{figure}[h]
\hspace{0cm}
\includegraphics[width=0.45\textwidth, height=0.35\textwidth]{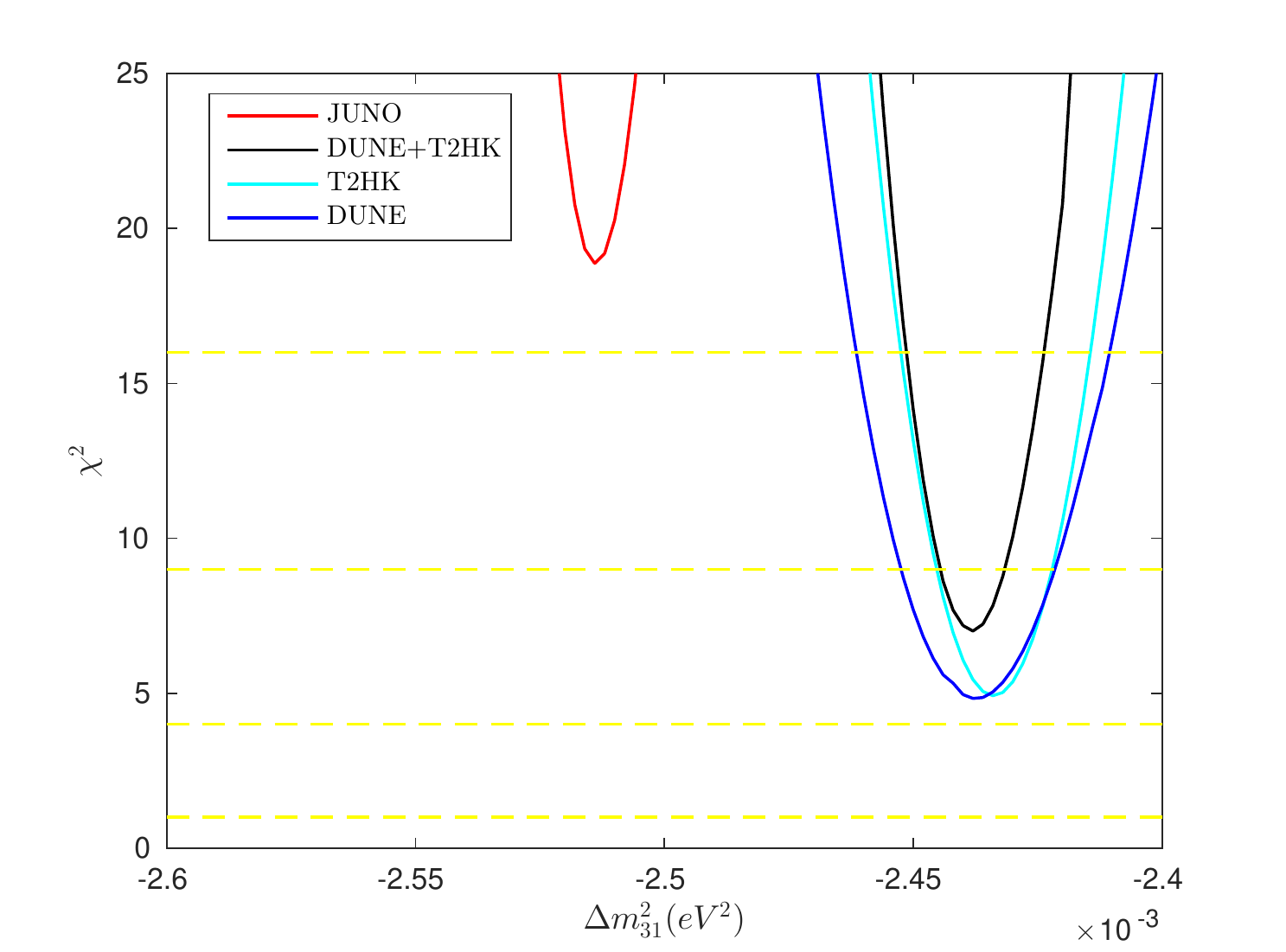}
\hspace{0cm}
\includegraphics[width=0.45\textwidth, height=0.35\textwidth]{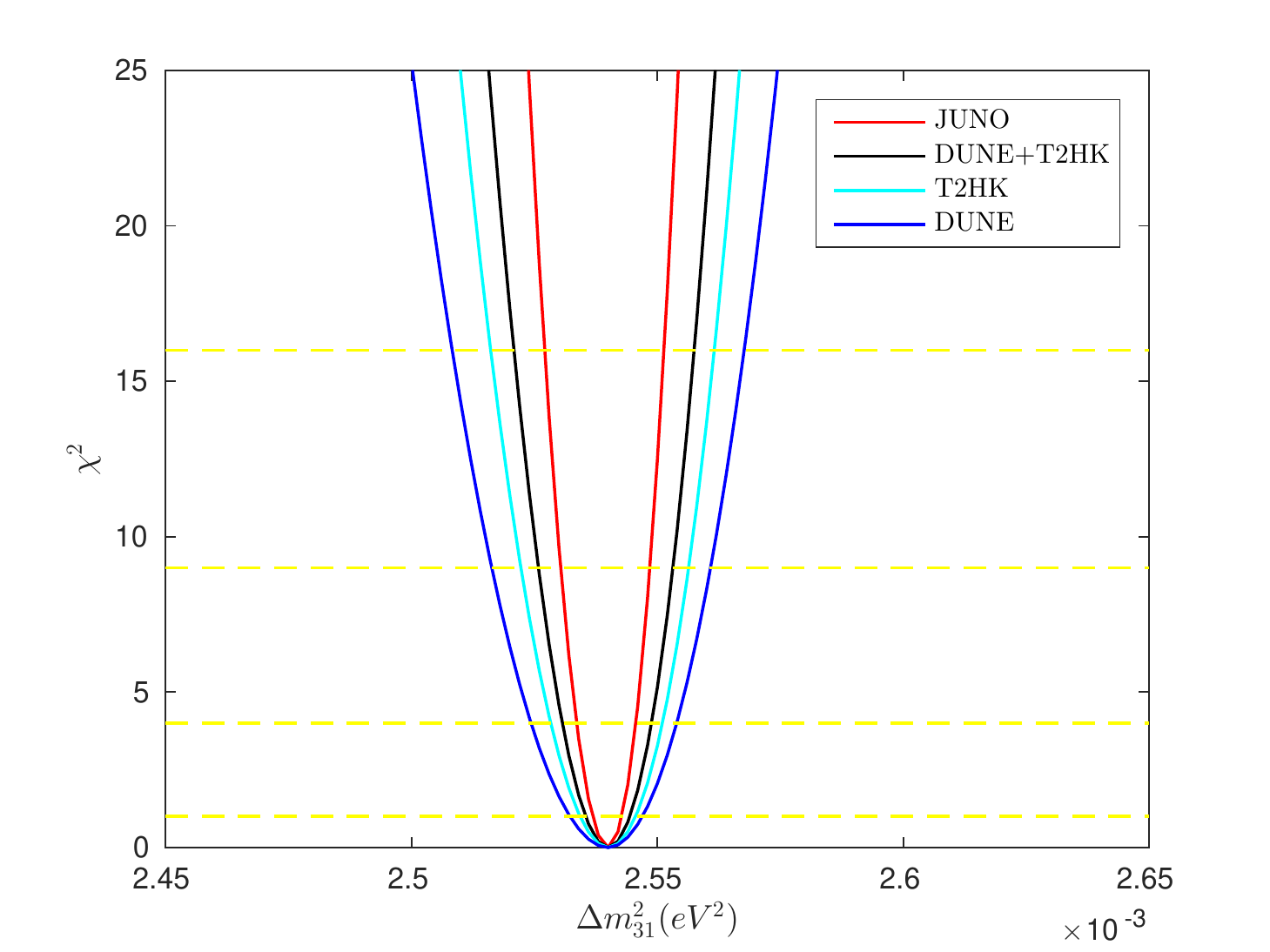}
\caption[...]{
Chi-squared vs. $\Delta m^2 _{31}$ assuming normal ordering as the true ordering. Red, black, cyan and blue curves show
chi-squared vs. $\Delta m^2 _{31}$ for JUNO, DUNE+T2HK, T2HK and DUNE, respectively, for 10 years of data taking. As it can be seen from the plot, In the left panel JUNO can exclude $\Delta m^2 _{13}<0$ at 4$\sigma$ C.L. while DUNE, T2HK and their combination exclude $\Delta m^2 _{13}<0$ at 2$\sigma$ C.L.. In the right panel, we can see that JUNE can determine the $\mid \Delta m^2 _{31} \mid$ better than T2HK. DUNE and their combination. The yellow dashed lines show 1$\sigma$, 2$\sigma$, 3$\sigma$ and 4$\sigma$ lines.
\label{fig2}
}
\end{figure}

Using the characteristics of DUNE, T2HK and JUNO, we show the results for the
analysis of measurement of $\Delta m^2 _{31}$ in the presence of NSI.
Fig.~\ref{fig2} indicates chi-squared vs. $\Delta m^2 _{31}$ assuming Normal Ordering (NO) as the true ordering. As can be seen from the figure,
$\Delta m^2 _{31}<0$ is excluded at 2$\sigma$ by DUNE, T2HK and their combination and at more than 4$\sigma$ by JUNO. As it can be seen in Fig.~\ref{fig2}-b, JUNO can determine $\Delta m^2 _{31}>0$ better than DUNE, T2HK and the combination of DUNE and T2HK.

\begin{figure}[h]
\hspace{0cm}
\includegraphics[width=0.55\textwidth, height=0.4\textwidth]{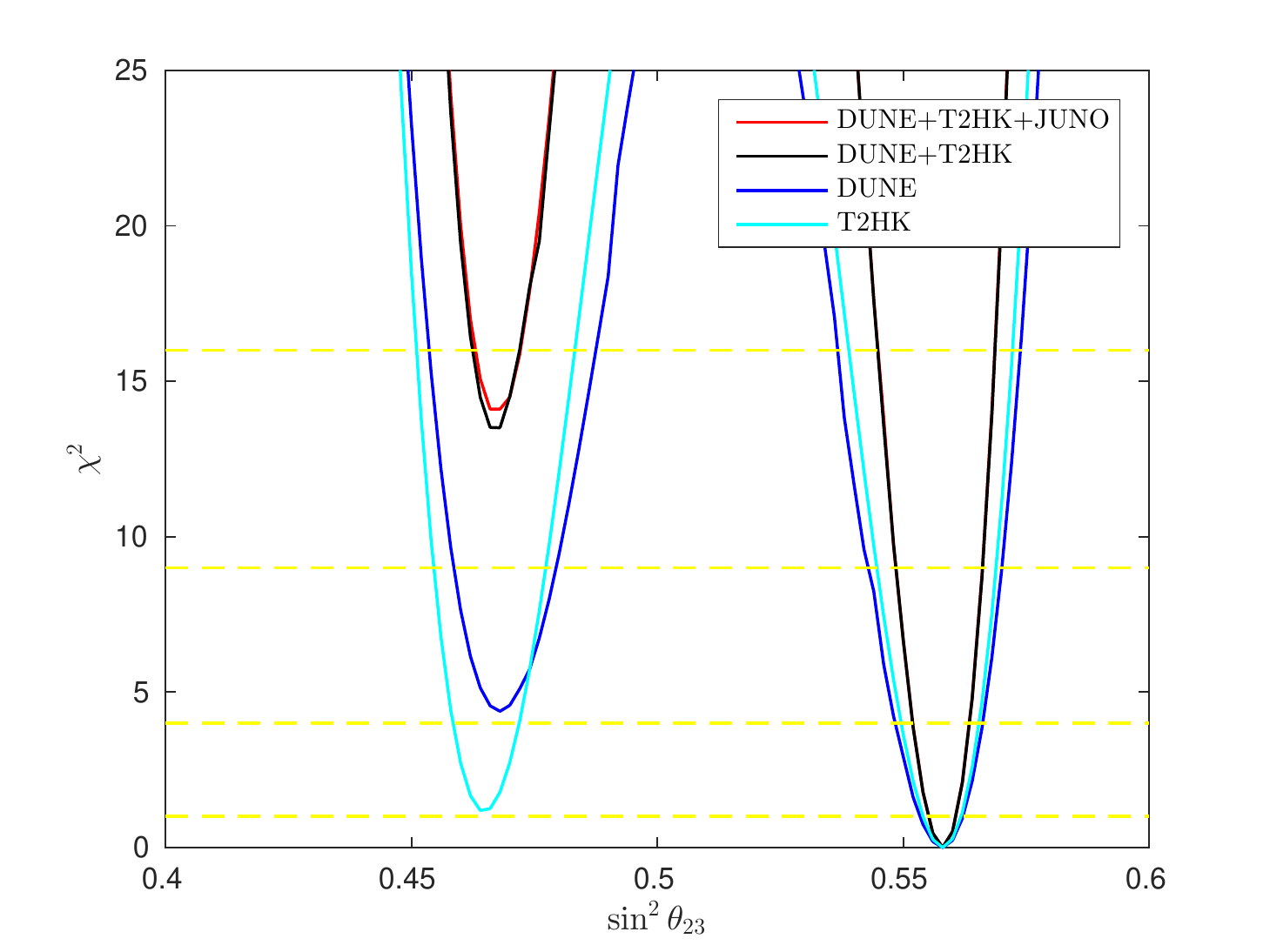}
\caption[...]{
Chi-squared vs. $\sin^2\theta_{23}$ assuming 10 years of data taking. The black, blue and the red curves indicate T2HK and DUNE and the combination of T2HK, DUNE and JUNO,
assuming $\sin^2 \theta_{23}=0.56$ \cite{Esteban:2020cvm}. As can be seen, The combination of the experiments has a strong sensitivity to the determination of the octant of $\theta_{23}$. Including JUNO increases the sensitivity but does not have a strong impact on the measurement of $\theta_{23}$.
\label{t23}
}
\end{figure}

Due to the importance of CP-violation phase and determination of the octant of $\theta_{23}$, we study how these future experiments can determine these parameters. Fig. \ref{t23} show the sensitivity of these experiments to the determination of $\theta_{23}$. We have set the true value of $\sin^2 \theta_{23}=0.56$ as given in \cite{Esteban:2020cvm}. As can be seen from the plot, DUNE and T2HK cannot determine the octant independently. Combining DUNE and T2HK data can determine the octant of $\theta_{23}$ as demonstrated. Although adding JUNO can improve the sensitivity of the determination of the octant ( one unit of chi-squared ), the combination of these experiments excludes the wrong octant at 4$\sigma$ C.L..

\begin{figure}[h]
\hspace{0cm}
\includegraphics[width=0.55\textwidth, height=0.4\textwidth]{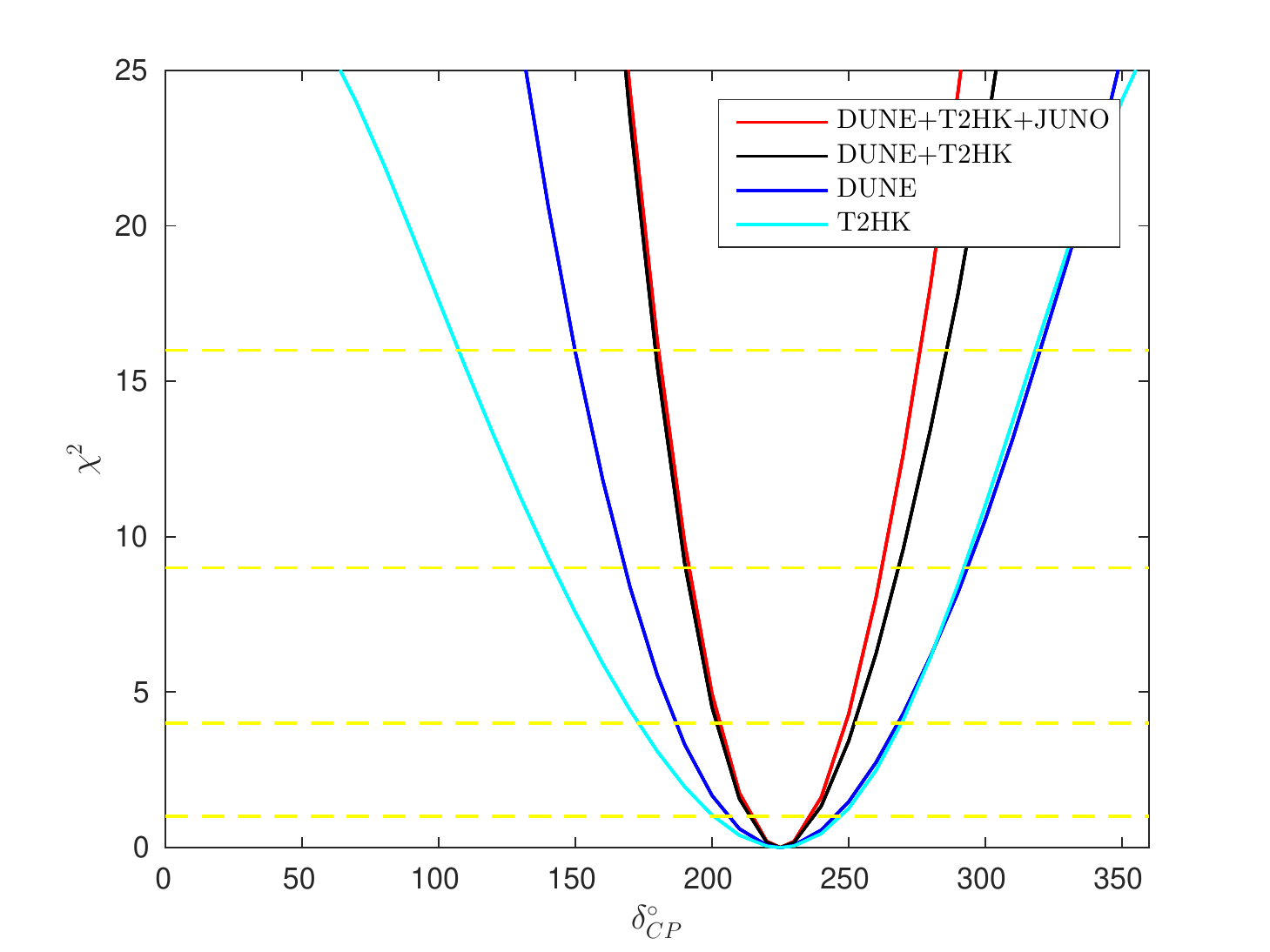}
\caption[...]{
Chi-squared vs. $\delta _{CP}$ assuming 10 years of data taking. The red, black, blue and the cyan curves corresponds to the combination of T2HK, DUNE and JUNO, the combination of T2HK and DUNE, DUNE and T2HK.
We assumed $\delta_{CP}=225^{\circ}$ \cite{Esteban:2020cvm}.
\label{delta}
}
\end{figure}

Fig.~\ref{delta} indicates chi-squared vs. $\delta_{CP}$ for T2HK, DUNE, the combination of T2HK and DUNE and for the combination of all three experiments,
assuming $\delta_{CP}=225^{\circ}$ as the true value \cite{Esteban:2020cvm}. As can be seen, T2HK and DUNE can exclude no CP-violation case at more than 1$\sigma$ and 2$\sigma$, respectively. Combining T2HK, DUNE and JUNO data can exclude the no CP-violation case at more than
4$\sigma$ C.L..

\begin{figure}[h]
\hspace{0cm}
\includegraphics[width=0.42\textwidth, height=0.32\textwidth]{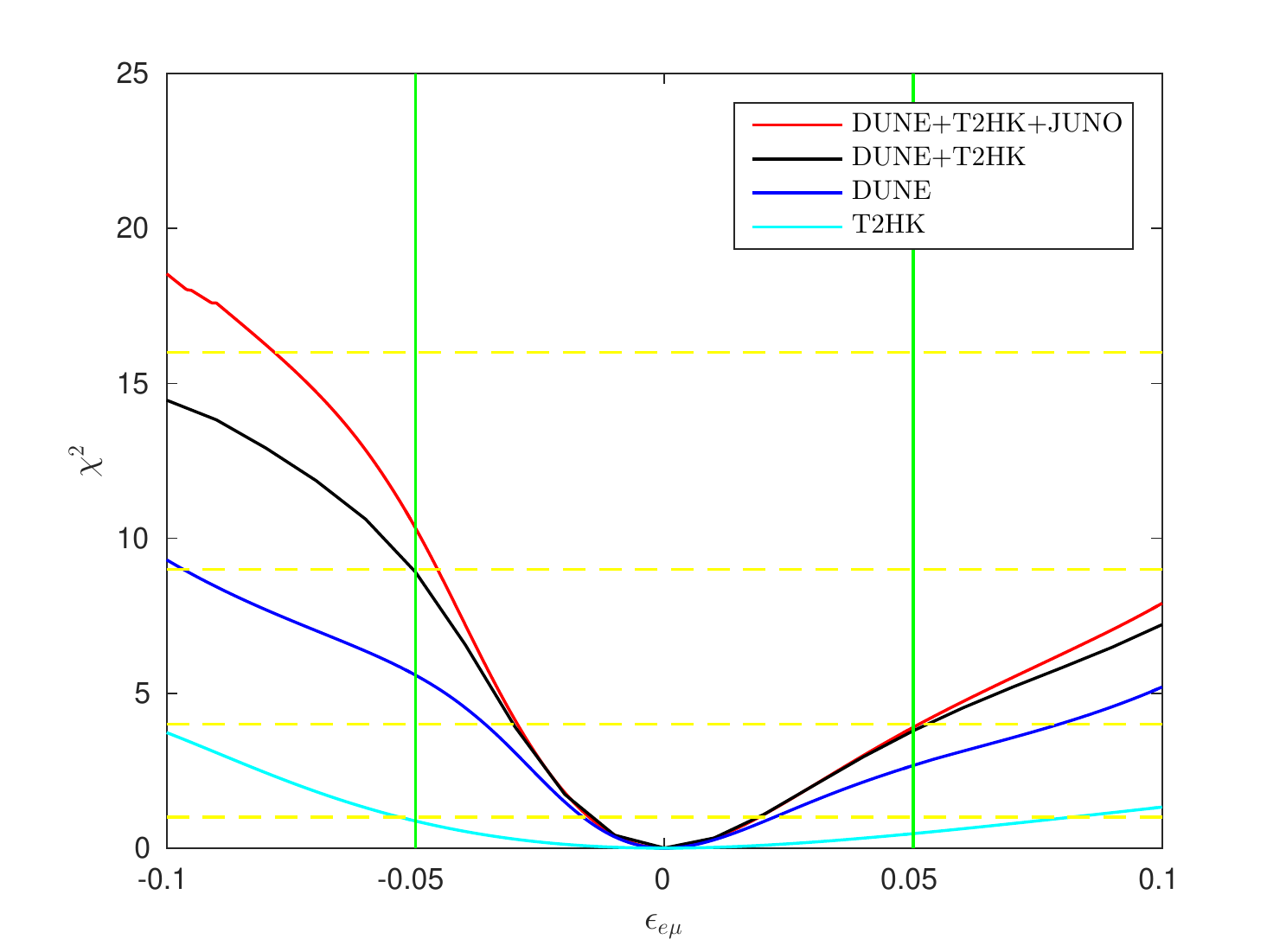}
\hspace{0cm}
\includegraphics[width=0.42\textwidth, height=0.32\textwidth]{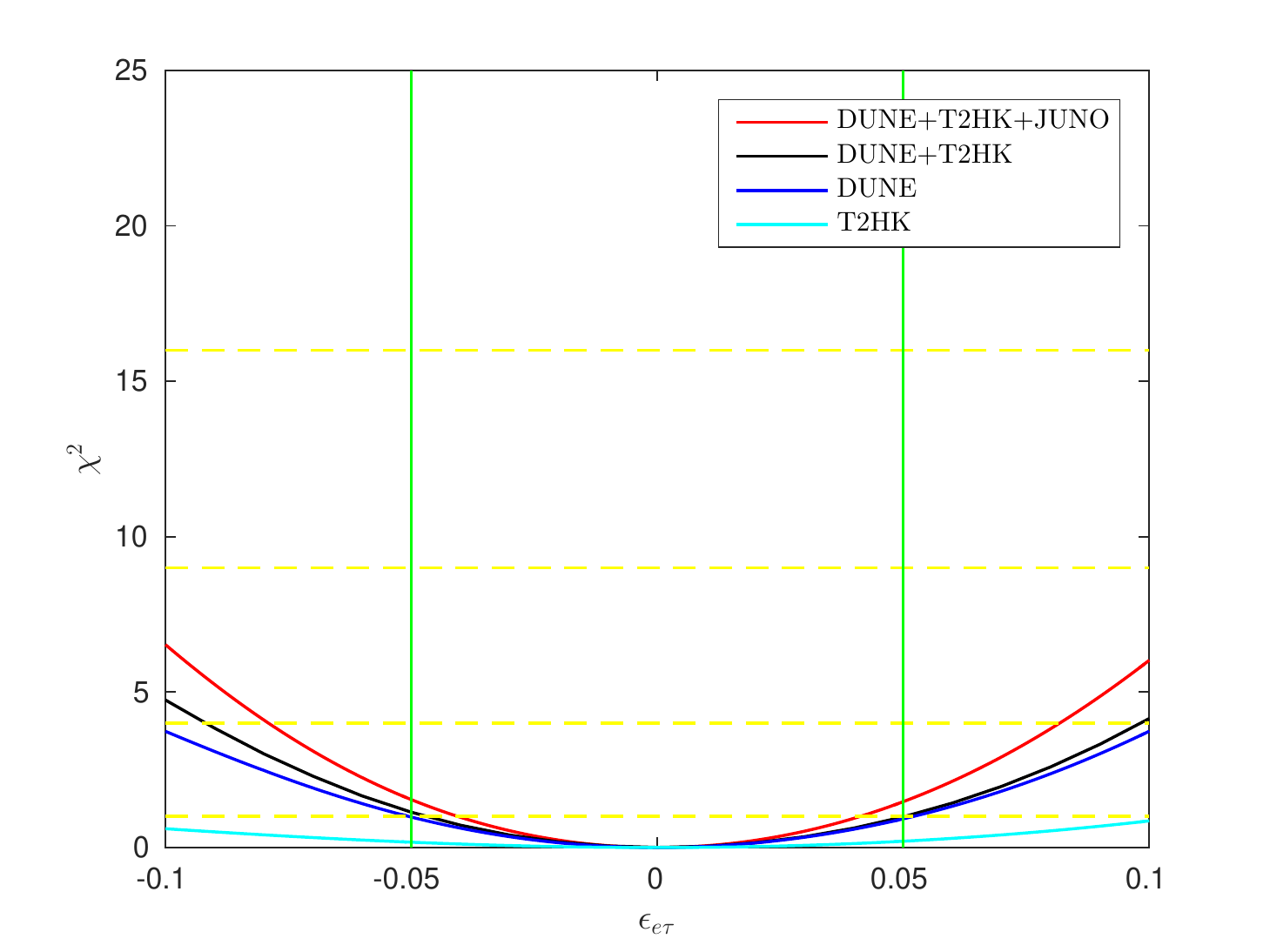}
\hspace{0cm}
\includegraphics[width=0.42\textwidth, height=0.32\textwidth]{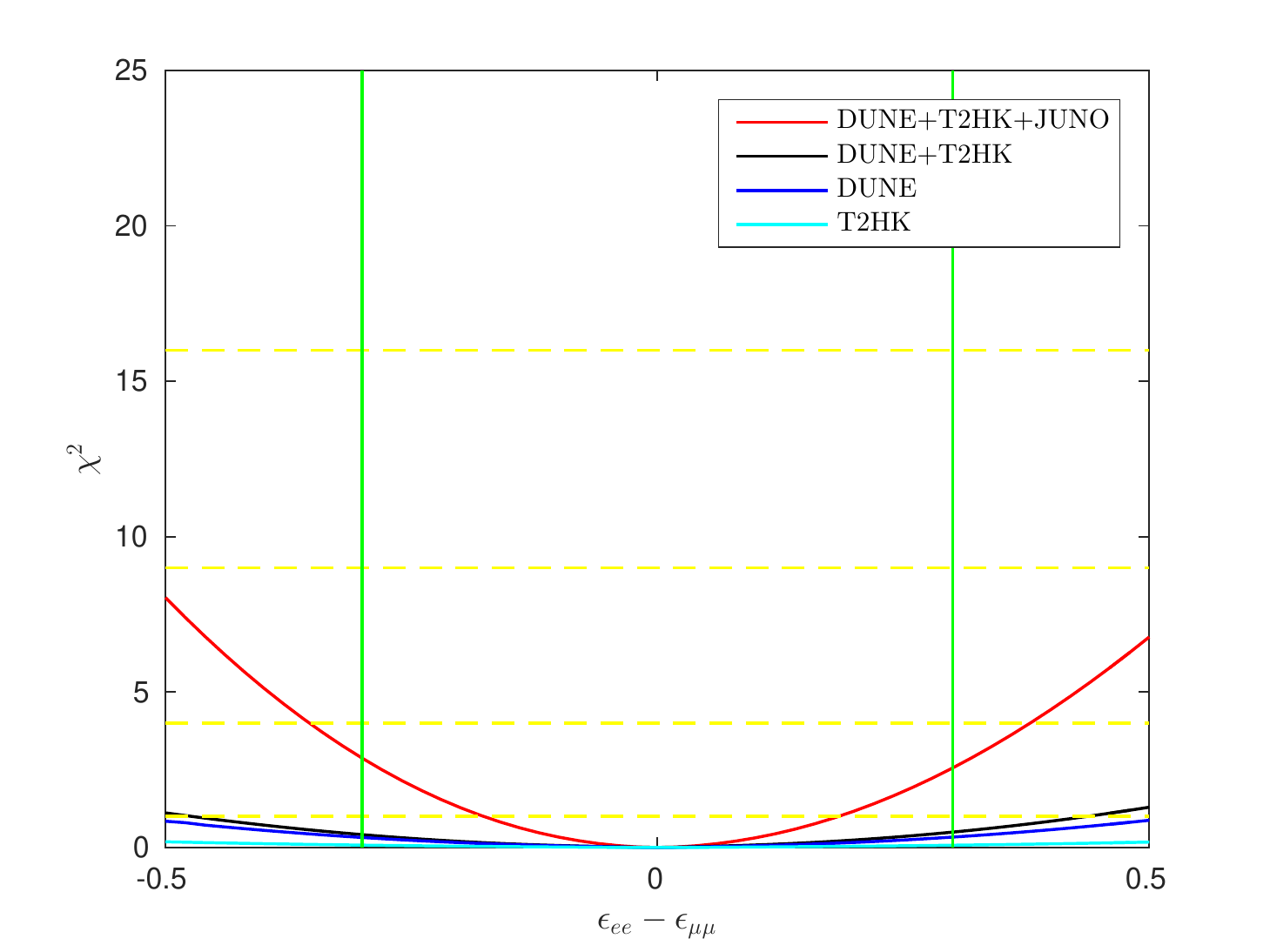}
\caption[...]{Chi-squared vs. $\epsilon_{e \mu}$, $\epsilon_{e \tau}$ and $\epsilon_{e e} -\epsilon_{\mu \mu} $. The red, black, blue and cyan curves correspond to the combined data of all experiments, the combination of T2HK and DUNE, DUNE and T2HK, respectively. As it can be seen from the plot,
DUNE is sensitive to the measurement of $\epsilon_{e \mu}$ and $\epsilon_{e \tau}$ while T2HK has a weak sensitivity to it. Considering the combination of these three experiments can increase the sensitivity to the measurement of $\epsilon_{e \mu}$ and $\epsilon_{e \tau}$.
In the lower panel, it can be seen that the combination of these experiments increases the sensitivity to the determination of $\epsilon_{e e} -\epsilon_{\mu \mu} $ while DUNE, T2HK and their combination are not sensitive to this parameter.
The green vertical lines indicate the current constraints at 1$\sigma$ C.L. \cite{Esteban:2019lfo}. The yellow dashed horizontal lines show the 1$\sigma$, 2$\sigma$, 3$\sigma$ and 4$\sigma$ region. } \label{fig4}
\end{figure}

\begin{figure}[h]
\hspace{0cm}
\includegraphics[width=0.42\textwidth, height=0.32\textwidth]{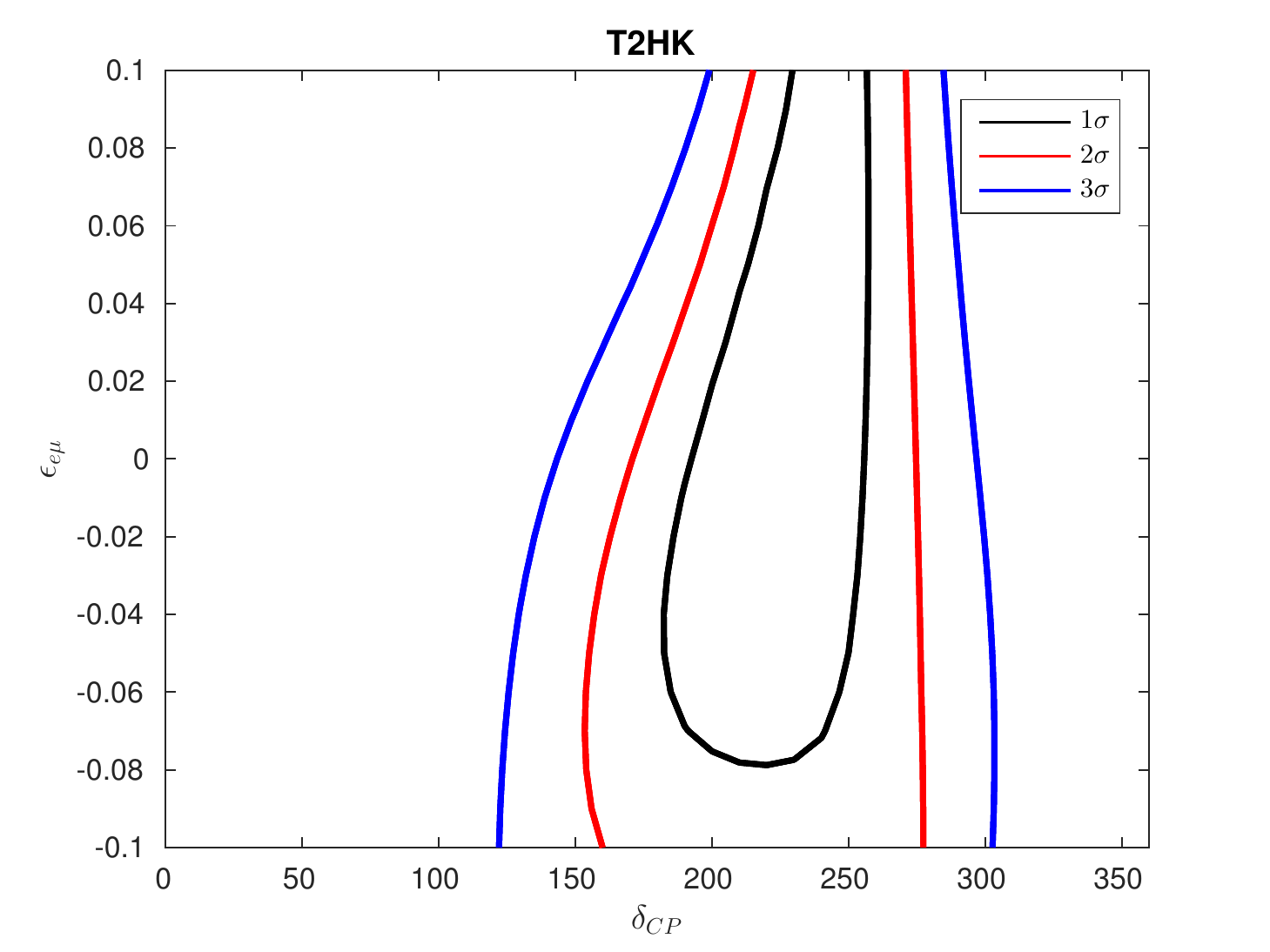}
\hspace{0cm}
\includegraphics[width=0.42\textwidth, height=0.32\textwidth]{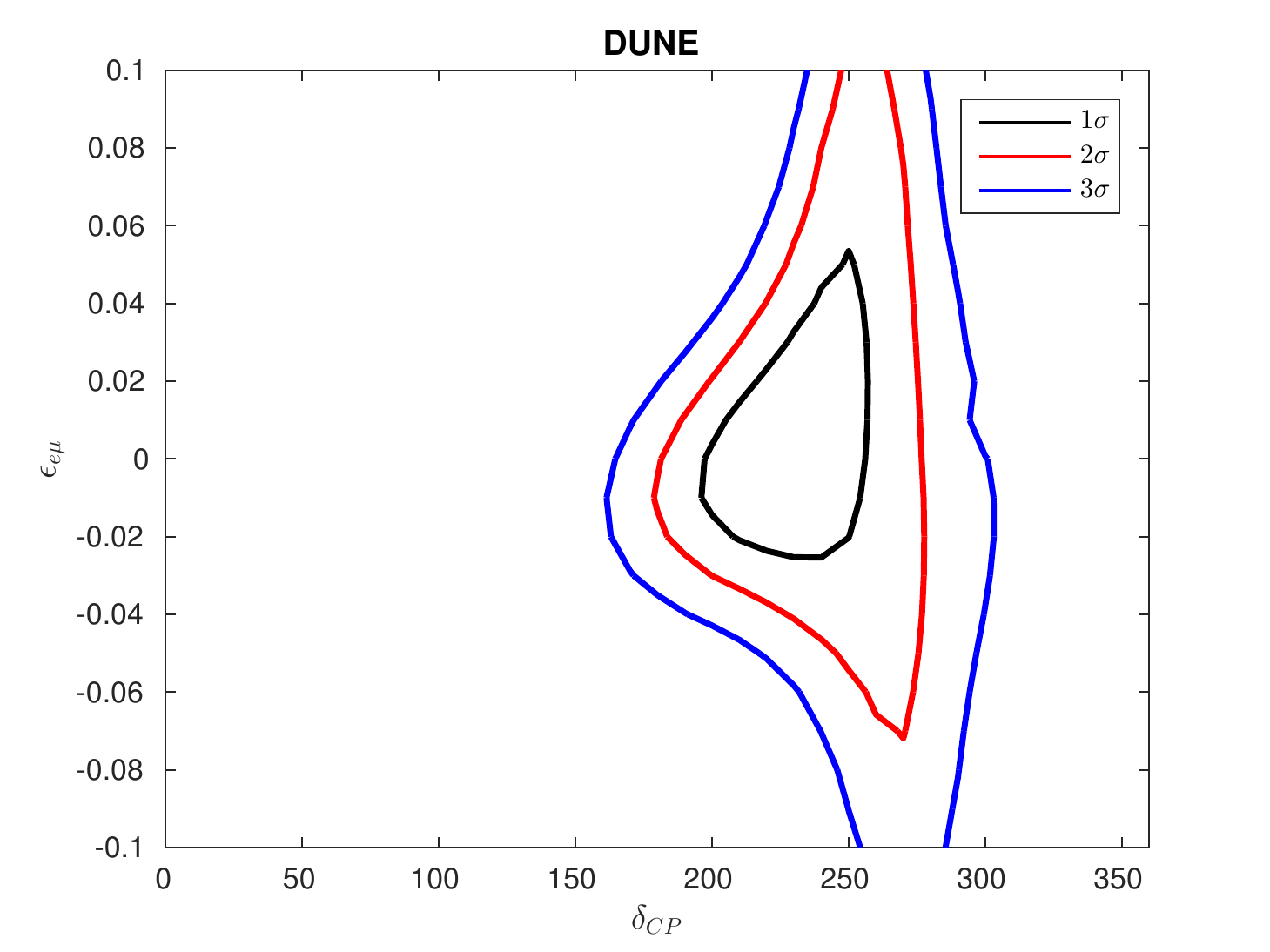}
\hspace{0cm}
\includegraphics[width=0.42\textwidth, height=0.32\textwidth]{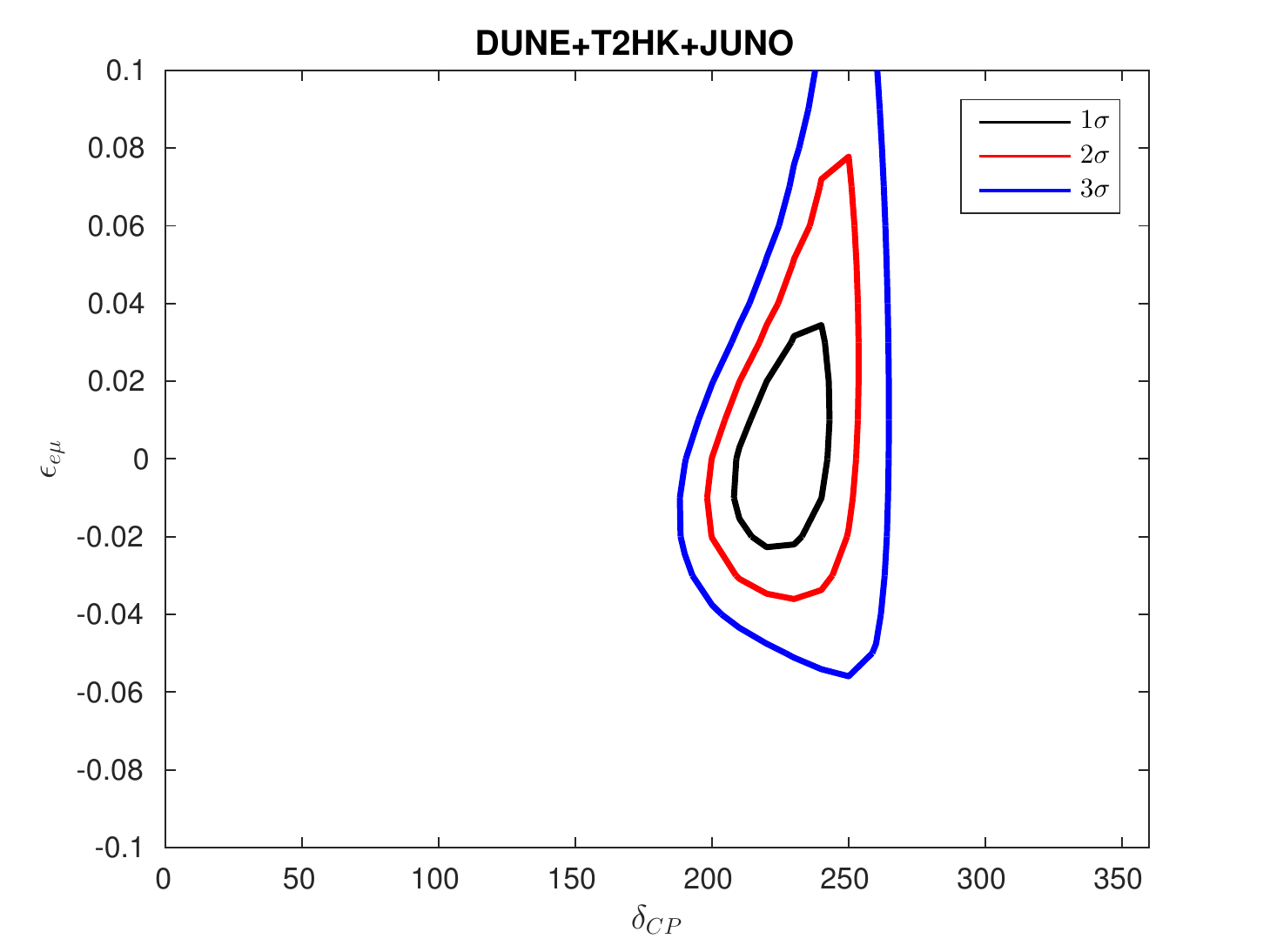}
\caption[...]{ $\epsilon_{e \mu}$ vs. $\delta_{CP}$ plotted for T2HK, DUNE and the combination of T2HK, DUNE and JUNO experiments.
As   can be seen, DUNE is sensitive to the simultaneous measurement of $\epsilon_{e \mu}$ vs. $\delta_{CP}$. Although DUNE and T2HK are sensitive to $\delta_{CP}$, they have a weak sensitivity to $\epsilon_{e \mu}$.  Combining T2HK, DUNE and JUNO can increase the sensitivity of the simultaneous measurement of $\epsilon_{e \mu}$ and $\delta_{CP}$ after 10 years of data taking. This combination improves $\delta_{CP}$ measurement slightly while improves the measurement of $\epsilon_{e \mu}$ strongly.} \label{sim}
\end{figure}

\begin{figure}[h]
\hspace{0cm}
\includegraphics[width=0.42\textwidth, height=0.32\textwidth]{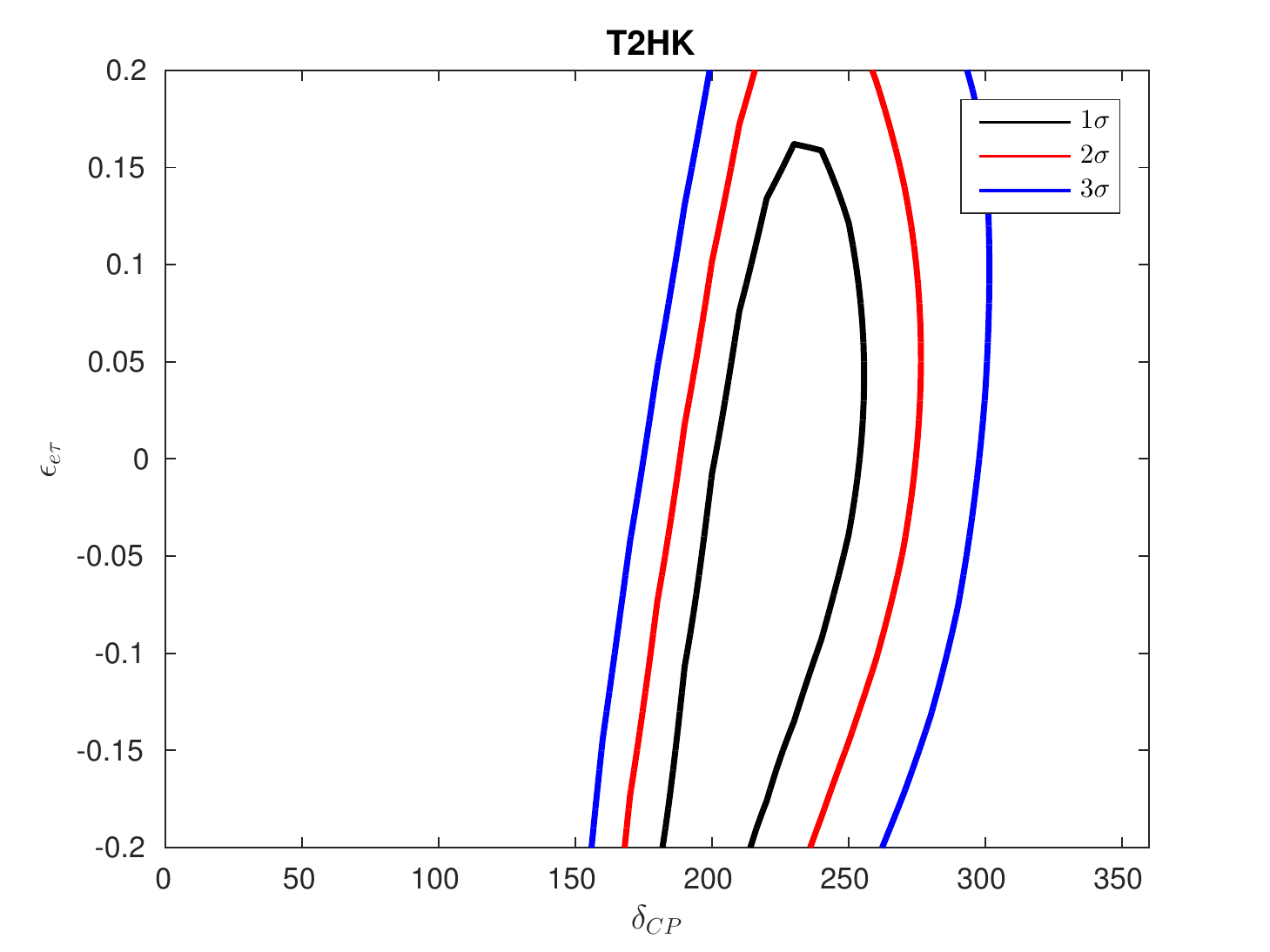}
\hspace{0cm}
\includegraphics[width=0.42\textwidth, height=0.32\textwidth]{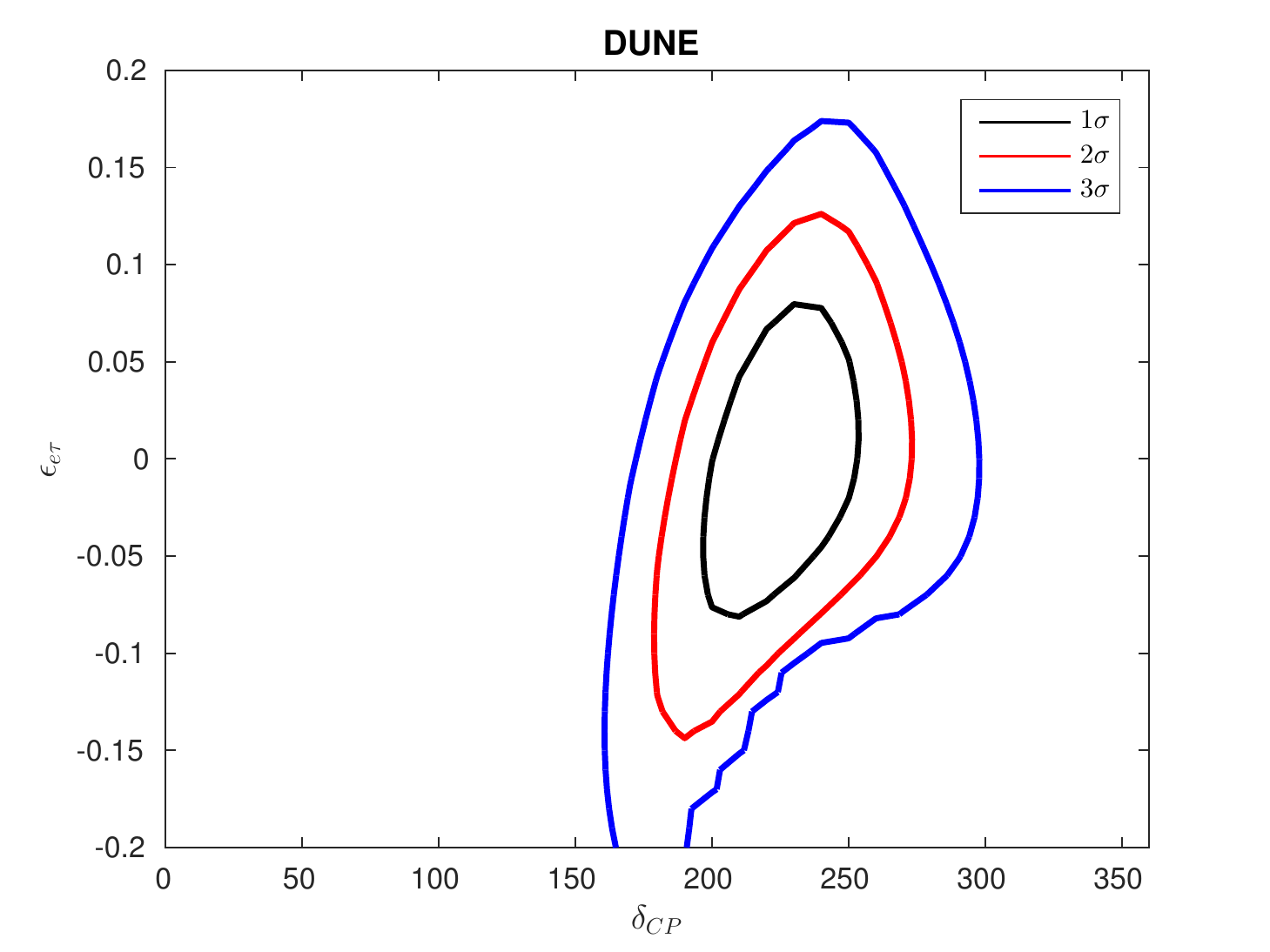}
\hspace{0cm}
\includegraphics[width=0.42\textwidth, height=0.32\textwidth]{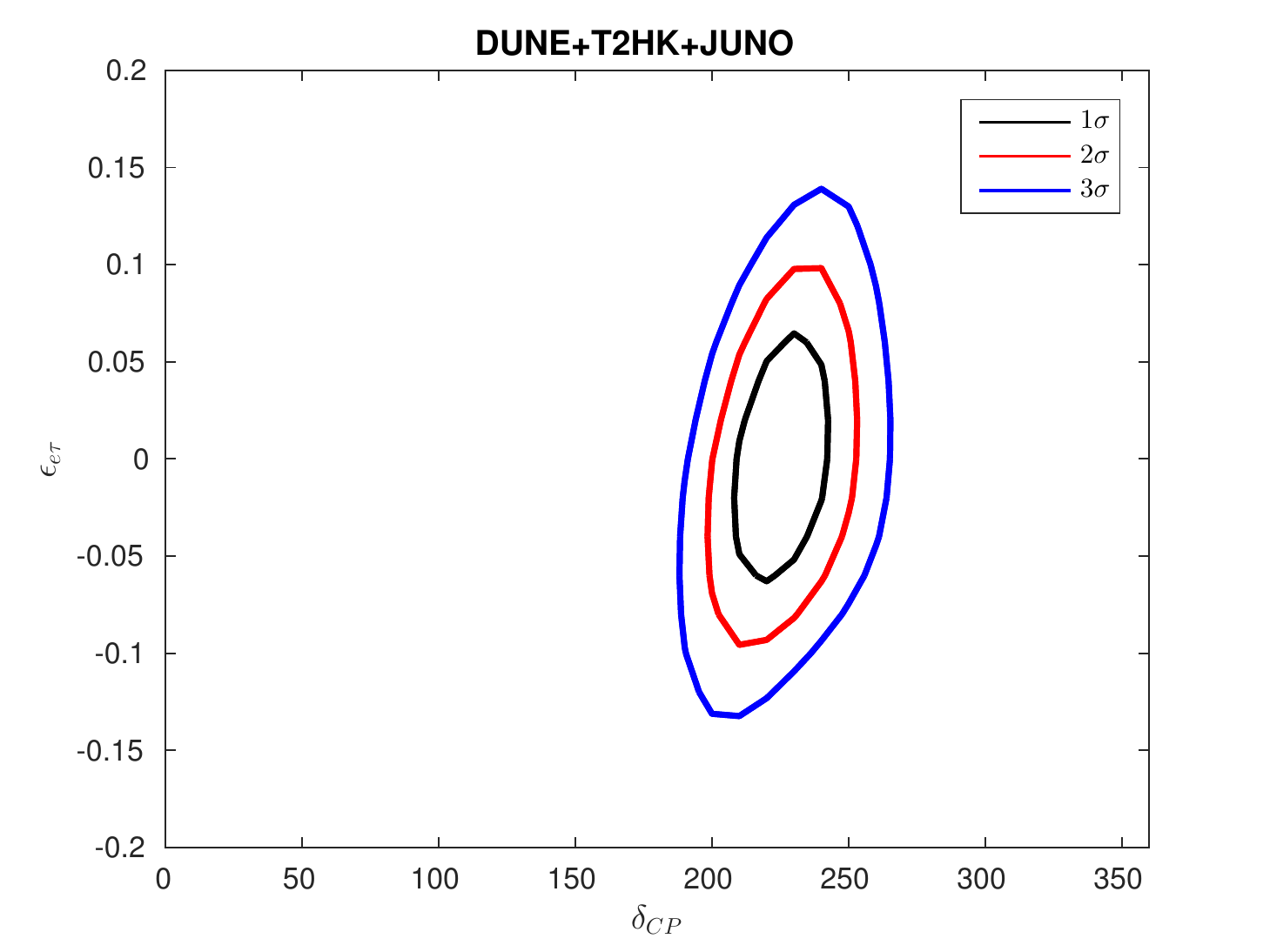}
\caption[...]{$\epsilon_{e \tau}$ vs. $\delta_{CP}$ plotted for T2HK, DUNE and the combination of T2HK, DUNE and JUNO experiments.
As   can be seen, DUNE is sensitive to the simultaneous measurement of $\epsilon_{e \tau}$ vs. $\delta_{CP}$. Although DUNE and T2HK are sensitive to $\delta_{CP}$, they have a weak sensitivity to $\epsilon_{e \tau}$.  Combining T2HK, DUNE and JUNO can increase the sensitivity of the simultaneous measurement of $\epsilon_{e \tau}$ and $\delta_{CP}$ after 10 years of data taking. This combination improves $\delta_{CP}$ measurement slightly while improves the measurement of $\epsilon_{e \tau}$ strongly.}\label{sim2}
\end{figure}

\begin{figure}[h]
\hspace{0cm}
\includegraphics[width=0.42\textwidth, height=0.32\textwidth]{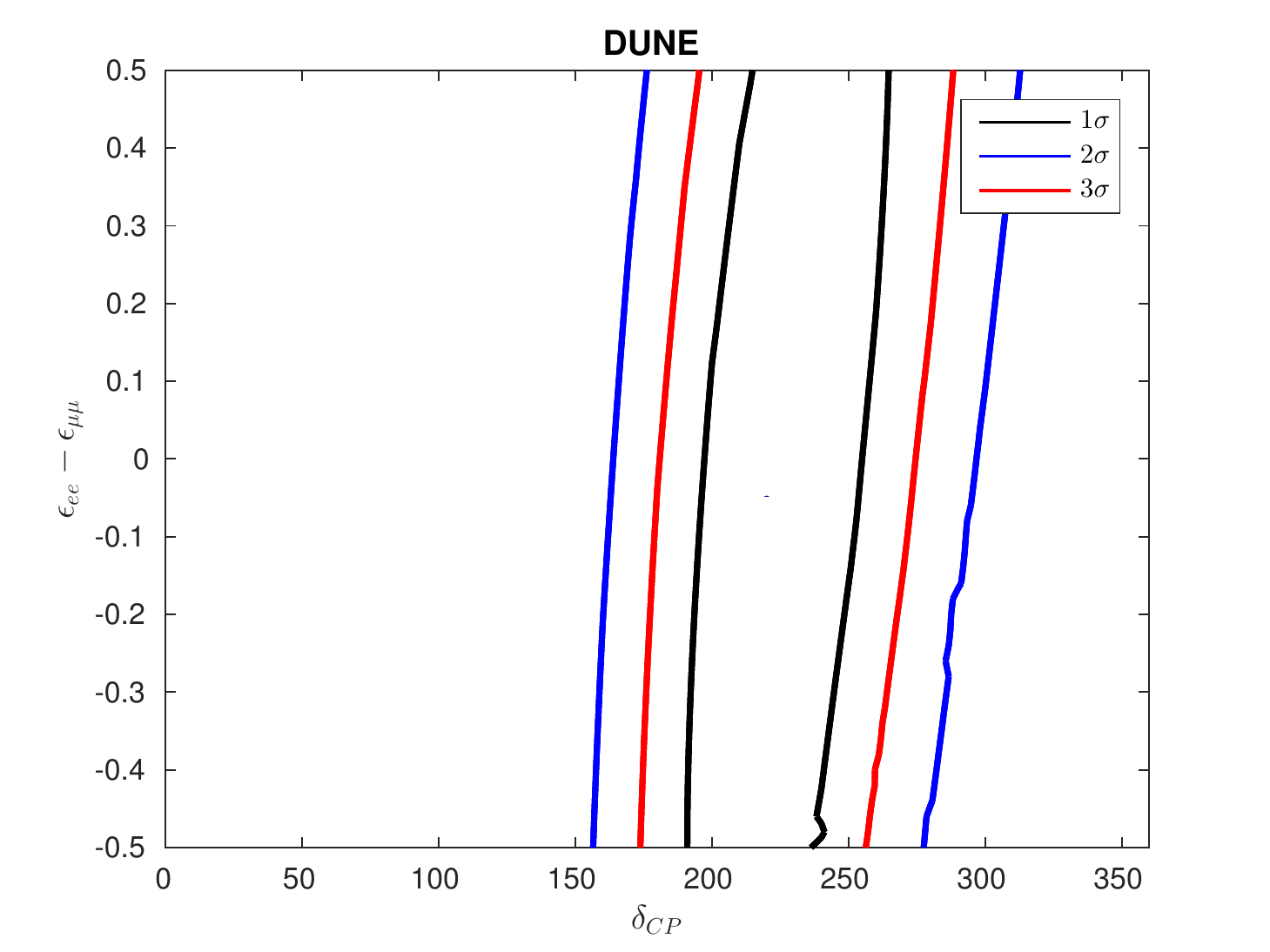}
\hspace{0cm}
\includegraphics[width=0.42\textwidth, height=0.32\textwidth]{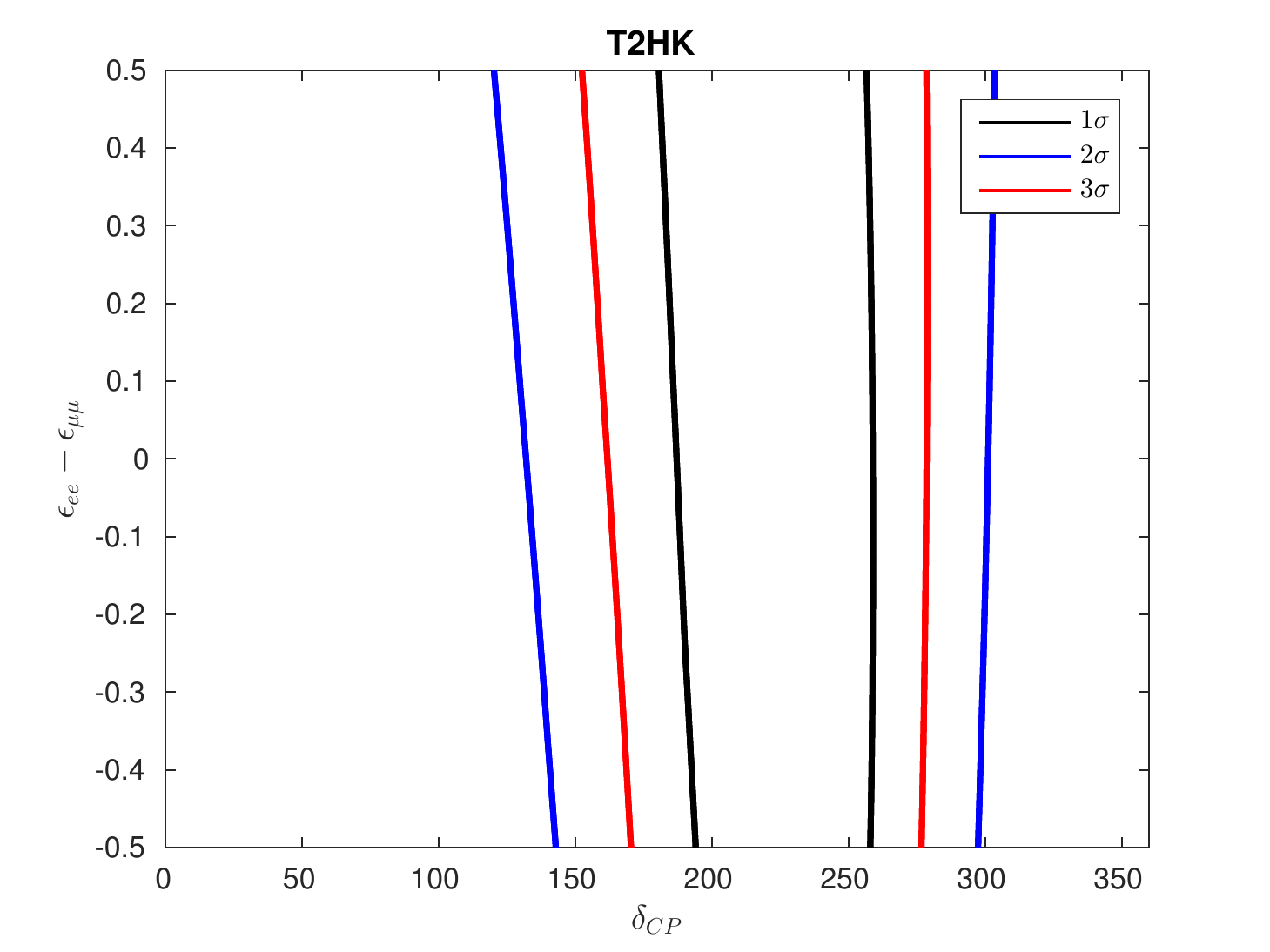}
\hspace{0cm}
\includegraphics[width=0.42\textwidth, height=0.32\textwidth]{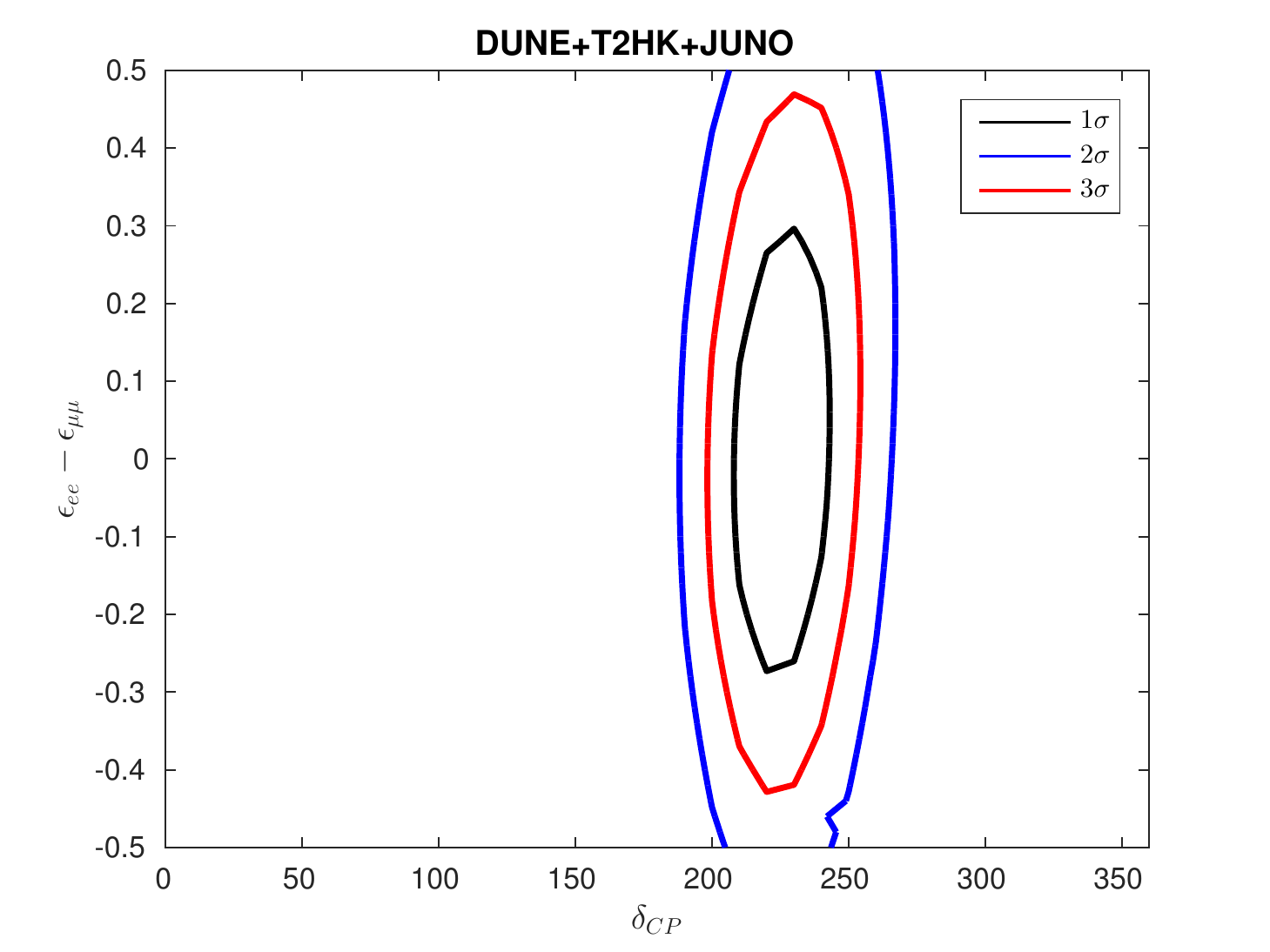}
\caption[...]{$\epsilon_{e e} - \epsilon_{\mu \mu}$ vs. $\delta_{CP}$ plotted for T2HK, DUNE and the combination of T2HK, DUNE and JUNO experiments.
As   can be seen, DUNE is sensitive to the simultaneous measurement of $\epsilon_{e e} - \epsilon_{\mu \mu}$ vs. $\delta_{CP}$. Although DUNE and T2HK are sensitive to $\delta_{CP}$, they have a weak sensitivity to $\epsilon_{e e} - \epsilon_{\mu \mu}$.  Combining T2HK, DUNE and JUNO can increase the sensitivity of the simultaneous measurement of $\epsilon_{e e} - \epsilon_{\mu \mu}$ and $\delta_{CP}$ after 10 years of data taking. This combination improves $\delta_{CP}$ measurement slightly while improves the measurement of $\epsilon_{e e} - \epsilon_{\mu \mu}$ strongly.}\label{sim3}
\end{figure}

Since the matter effect is more important for DUNE than for T2HK, (because of the larger energy and the longer baseline of DUNE), DUNE is more sensitive to the determination of the NSI parameters. We assume that the data are consistent with the SM and the NO, and plot the chi-squared vs. $\epsilon_{e \mu}$, $\epsilon_{e \mu}$ and $\epsilon_{e e} -\epsilon_{\mu \mu} $ for the following four cases: considering T2HK data, considering DUNE data, considering the combination of DUNE and T2HK data and considering the combined data of T2HK, DUNE and JUNO. The results are shown in Fig.~\ref{fig4}. As it can be seen from the figure, while
DUNE is more sensitive to the measurement of $\epsilon_{e \mu}$ and $\epsilon_{e \tau}$ than T2HK, considering the combination of T2HK, DUNE and JUNO data increase the accuracy to determine $\epsilon_{e \mu}$ and $\epsilon_{e \tau}$. Since JUNO reduces the uncertainties of the oscillation parameters, this combination works better.
As indicated in the lower panel of Fig.~\ref{fig4}, DUNE and T2HK cannot determine $\epsilon_{e e} -\epsilon_{\mu \mu} $. Moreover, the combined analysis of DUNE and T2HK data shows that their combination is not sensitive to the determination of $\epsilon_{e e} -\epsilon_{\mu \mu} $. However, combining the data of DUNE and T2HK experiments while including JUNO data, is interestingly sensitive to the determination of this parameter. The vertical green lines indicate the current constraints \cite{Esteban:2019lfo}. DUNE can constrain $\epsilon_{e \mu}$ independently more stringent than the current constraints. Combining the data of DUNE, T2HK and JUNO can determine $\epsilon_{e \mu}$, $\epsilon_{e \tau}$ and $\epsilon_{e e} -\epsilon_{\mu \mu} $ more stringent than the current constraints.

We show the results for the
analysis of the simultaneous measurements of $\epsilon_{e \mu}$ and $\delta_{CP}$ in Fig.~\ref{sim}, considering DUNE, T2HK and the combination of T2HK, DUNE and JUNO.
Although DUNE and T2HK can constrain $\delta_{CP}$, they do not have a strong sensitivity to $\epsilon_{e \mu}$.  As demonstrated, combining T2HK, DUNE and JUNO can increase the sensitivity of the simultaneous measurement of $\epsilon_{e \mu}$ and $\delta_{CP}$. In more detail, including JUNO improves $\delta_{CP}$ slightly while improving strongly the determination of $\epsilon_{e \mu}$.
 Fig.~\ref{sim2} shows the results of combining T2HK, DUNE and JUNO to measure simultaneously $\epsilon_{e \tau}$ and $\delta_{CP}$. In the same way, the combination of data of these experiments can increase the accuracy of the
simultaneous measurements of $\epsilon_{e \tau}$ and $\delta_{CP}$.

Finally, we discuss simultaneous measurement of $\epsilon_{e e} - \epsilon_{\mu \mu}$ and $\delta_{CP}$ for T2HK, DUNE and the combination of T2HK, DUNE and JUNO experiments. Although DUNE and T2HK are not sensitive to the simultaneous measurement of $\epsilon_{e e} - \epsilon_{\mu \mu}$ independently, combining T2HK, DUNE and JUNO has a strong sensitivity to it and can increase the sensitivity to the determination of $\delta_{CP}$.

\section{Summary \label{sum}}

In this paper, we explored the impact of NSI on future long-baseline neutrino experiments DUNE and T2HK in addition to future reactor experiment JUNO.
We investigated the potential of these experiments to the determination of the oscillation parameters in the presence of NSI. Moreover, we studied the possibility to constrain NSI parameters using these future neutrino experiments.

Having relatively low energy, the reactor experiment, JUNO is not sensitive to the matter effects. JUNO is going to
provide the opportunity for the precise measurement of $\Delta m^2 _{21}$, $\Delta m^2 _{21}$, $\theta_{12}$. We assumed the same NSI couplings for electron, up and down quarks.

We performed detailed numerical simulations using GLoBES and studied the measurements of standard neutrino oscillation parameters in the presence of NSI using combined data from future experiments, DUNE, T2HK and JUNO. We have assumed 10 years of data taking for all of the experiments.
DUNE, T2HK and their combination of data are sensitive to the determination of mass ordering at 2$\sigma$ C.L. while JUNO can determine the mass ordering at more than 4$\sigma$ as indicated in Fig.~\ref{fig2}. DUNE and T2HK are not sensitive to the octant of $\theta_{23}$, separately while our results show that the combined data of these three experiments can determine the octant at about 4$\sigma$.
As it is indicated in Fig.~\ref{delta}, DUNE and T2HK can determine $\delta_{CP}$ separately; However, the combination of data from DUNE, T2HK and JUNO can increase the sensitivity to the determination of $\delta_{CP}$.
As can be seen, the case of no CP-violation can be excluded at 2$\sigma$ using DUNE and T2HK. The combination of data of these two experiments can exclude it at 4$\sigma$ C.L. (Fig.\ref{delta}).

While DUNE is sensitive to the determination of $\epsilon_{e \mu}$ and $\epsilon_{e \tau}$ and is going to constrain these parameters stronger than the current constraints, T2HK has a weak sensitivity to it. We have shown that considering the combination of DUNE, T2HK and JUNO, can determine $\epsilon_{e \mu}$ and $\epsilon_{e \tau}$ much better than DUNE. Moreover, DUNE, T2HK and the combination of them are not sensitive to $\epsilon_{e e} - \epsilon_{\mu \mu}$. We have shown that considering the combination of DUNE, T2HK and JUNO is sensitive to the determination of $\epsilon_{e e} - \epsilon_{\mu \mu}$ and can constrain it more stringent than the current constraints (Fig.~\ref{fig4}).

In addition, we have performed a combined analysis to study how combined data from DUNE, T2HK and JUNO is sensitive to the simultaneous measurement of $\epsilon_{e \mu}$ and $\delta_{CP}$ (Fig.~\ref{sim}). In the same way, we studied how the combined analysis of DUNE, T2HK and JUNO data can determine the simultaneous measurement of $\epsilon_{e \tau}$ and $\delta_{CP}$ as well as the simultaneous measurement of $\epsilon_{e e} - \epsilon_{\mu \mu}$ and $\delta_{CP}$ (Fig.~\ref{sim2} and Fig.~\ref{sim3}). In more detail, DUNE and T2HK are sensitive to $\delta_{CP}$ while having a weak sensitivity to NSI parameters. Combining these experiments with JUNO improves slightly the sensitivity to $\delta_{CP}$ while increases the sensitivity to NSI parameters significantly. 


Overall the combined analysis of future long-baseline experiments DUNE and T2HK, as well as the future reactor experiment JUNO, is going to determine the standard oscillation parameters in the presence of NSI. Besides, the combination of these experiments can constrain the NSI parameters with higher precision.

 \subsection*{Acknowledgments}
This project has received funding from the European Union's Horizon 2020 research and innovation programme under the Marie Sk\l{}odowska-Curie grant agreement No.~674896 and No.~690575.  P.B thanks Iran Science Elites Federation Grant No. 11131.



\begin{thebibliography}{99}
\bibitem{Wolf}
L. Wolfenstein, Neutrino oscillations in matter, Phys. Rev. D
17, 2369 (1978).


\bibitem{ns1}
P. Minkowski, 
 at a Rate of One Out of 109 Muon Decays , Phys.Lett. B67 (1977)
421 428.

\bibitem{ns2}
 T. Yanagida, HORIZONTAL SYMMETRY AND MASSES OF NEUTRINOS, Conf.Proc.
C7902131 (1979) 95  99.

\bibitem{ns3}
  R. N. Mohapatra and G. Senjanovic, Neutrino Mass and Spontaneous Parity Violation,
Phys.Rev.Lett. 44 (1980) 912.

\bibitem{ns4}
 M. Gell-Mann, P. Ramond, and R. Slansky, Complex Spinors and Unified Theories,
Conf.Proc. C790927 (1979) 315 321, [arXiv:1306.4669].

\bibitem{ns5}
 J. Schechter and J. Valle, Neutrino Masses in SU(2) times U(1) Theories, Phys.Rev. D22
(1980) 2227.

\bibitem{ns6}
 G. Lazarides
 , Q. Shafi, and C. Wetterich, Proton Lifetime and Fermion Masses in an
SO(10) Model, Nucl.Phys. B181 (1981) 287 300.


\bibitem{ns7}
Y.~Farzan and M.~Tortola,
Front. in Phys. \textbf{6} (2018), 10
[arXiv:1710.09360 [hep-ph]].

\bibitem{mixing1}
 M. Gonzalez-Garcia, M. Maltoni, and T. Schwetz, Updated fit to three neutrino mixing:
status of leptonic CP violation, JHEP 1411 (2014) 052, [arXiv:1409.5439].

\bibitem{mixing2}
 F. Capozzi, G. Fogli, E. Lisi, A. Marrone, D. Montanino, et al., Status of three-neutrino
oscillation parameters, circa 2013, Phys.Rev. D89 (2014) 093018, [arXiv:1312.2878].

\bibitem{mixing3}
 D. Forero, M. Tortola, and J. Valle, Neutrino oscillations retted, Phys.Rev. D90 (2014)
093006, [arXiv:1405.7540].



\bibitem{Sol}
M. M. Guzzo, A. Masiero, and S. T. Petcov, Phys. Lett. B260, 154 (1991).


\bibitem{sol1}
P. I. Krastev and S. T. Petcov, Phys. Lett. B299, 99 (1993).

\bibitem{sol2}
O. G. Miranda, M. A. Tortola, and J. W. F. Valle, JHEP 10, 008 (2006), hep-ph/0406280

\bibitem{sol3}
A. Bola~nos, O. G. Miranda, A. Palazzo, M. A. Tortola, and J. W. F. Valle, Phys. Rev. D79, 113012 (2009), 0812.4417

\bibitem{sol4}
A. Palazzo and J. W. F. Valle, Phys. Rev. D80, 091301 (2009), 0909.1535.

\bibitem{sol5}
F. J. Escrihuela, O. G. Miranda, M. A. Tortola, and J. W. F. Valle, Phys. Rev. D80, 105009 (2009), [Erratum: Phys.
Rev.D80,129908(2009)], 0907.2630.

\bibitem{Bakhti:2020hbz}
P.~Bakhti and M.~Rajaee,
Phys. Rev. D \textbf{102} (2020) no.3, 035024
[arXiv:2003.12984 [hep-ph]].


\bibitem{atm1}
 M. C. Gonzalez-Garcia, M. M. Guzzo, P. I. Krastev, H. Nunokawa, O. L. G. Peres, V. Pleitez, J. W. F. Valle, and
R. Zukanovich Funchal, Phys. Rev. Lett. 82, 3202 (1999), hep-ph/9809531.

\bibitem{atm2}
  N. Fornengo, M. C. Gonzalez-Garcia, and J. W. F. Valle, JHEP 07, 006 (2000), hep-ph/9906539.
\bibitem{atm3}
 N. Fornengo, M. Maltoni, R. Tomas, and J. W. F. Valle, Phys. Rev. D65, 013010 (2002), hep-ph/0108043.
\bibitem{atm4}
 P. Huber and J. W. F. Valle, Phys. Lett. B523, 151 (2001), hep-ph/0108193.
\bibitem{atm5}
 A. Friedland, C. Lunardini, and M. Maltoni, Phys. Rev. D70, 111301 (2004), hep-ph/0408264.
\bibitem{atm6}
 A. Friedland and C. Lunardini, Phys. Rev. D72, 053009 (2005), hep-ph/0506143.
\bibitem{atm7}
 O. Yasuda, Nucl. Phys. Proc. Suppl. 217, 220 (2011), 1011.6440.
\bibitem{atm8}
 M. C. Gonzalez-Garcia, M. Maltoni, and J. Salvado, JHEP 05, 075 (2011), 1103.4365.
\bibitem{atm9}
 A. Esmaili and A. Yu. Smirnov, JHEP 06, 026 (2013), 1304.1042.
\bibitem{atm10}
 S. Choubey and T. Ohlsson, Phys. Lett. B739, 357 (2014), 1410.0410.
\bibitem{atm11}
 I. Mocioiu and W. Wright, Nucl. Phys. B893, 376 (2015), 1410.6193.
\bibitem{atm12}
 S. Fukasawa and O. Yasuda (2015), 1503.08056.
\bibitem{atm13} 
S. Choubey, A. Ghosh, T. Ohlsson, and D. Tiwari (2015), 1507.02211.

\bibitem{ac1}
 A. Friedland and C. Lunardini, Phys. Rev. D74, 033012 (2006), hep-ph/0606101.
\bibitem{ac2}
 M. Blennow, T. Ohlsson, and J. Skrotzki, Phys. Lett. B660, 522 (2008), hep-ph/0702059.
\bibitem{ac3}
 A. Esteban-Pretel, J. W. F. Valle, and P. Huber, Phys. Lett. B668, 197 (2008), 0803.1790.
\bibitem{ac4}
 J. Kopp, P. A. N. Machado, and S. J. Parke, Phys. Rev. D82, 113002 (2010), 1009.0014.
\bibitem{ac5}
 P. Coloma, A. Donini, J. Lopez-Pavon, and H. Minakata, JHEP 08, 036 (2011), 1105.5936.
\bibitem{ac6}
 A. Friedland and I. M. Shoemaker (2012), 1207.6642.
\bibitem{ac7} 
J. A. B. Coelho, T. Kafka, W. A. Mann, J. Schneps, and O. Altinok, Phys. Rev. D86, 113015 (2012), 1209.3757.
\bibitem{ac8}
 P. Adamson et al. (MINOS), Phys. Rev. D88, 072011 (2013), 1303.5314.

\bibitem{Bakhti:2016prn}
P.~Bakhti and Y.~Farzan,
JHEP \textbf{07} (2016), 109
[arXiv:1602.07099 [hep-ph]].

\bibitem{Bakhti:2016gic}
P.~Bakhti, A.~N.~Khan and W.~Wang,
J. Phys. G \textbf{44} (2017) no.12, 125001
[arXiv:1607.00065 [hep-ph]].

\bibitem{nsi1}
A. N. Khan, D. W. McKay and F. Tahir, 
Phys. Rev. D 88, 113006 (2013), arXiv:1305.4350.
\bibitem{nsi2}
 A. N. Khan, D. W. McKay and F. Tahir, 
 Phys.Rev. D 90, 053008 (2014), arXiv:1407.4263.
\bibitem{nsi3}
 A. N. Khan, 
 Phys. Rev. D 93 (2016) no.9, 093019, arXiv:1605.09284.
\bibitem{nsi4}
 I. Girardi, D. Meloni and S. T. Petcov, 
Nucl. Phys. B 886 (2014) 31, arXiv:1405.0416.
\bibitem{nsi5}
 I. Girardi and D. Meloni, 
 Phys. Rev. D 90
(2014) no.7, 073011, arXiv:1403.5507.
\bibitem{nsi6}
 A. de Gouvea and K. J. Kelly, 
 Nucl. Phys. B 908 (2016) 318, arXiv:1511.05562.
\bibitem{nsi7}
 J. Liao, D. Marfatia and K. Whisnant, 
arXiv:1601.00927.
\bibitem{nsi8}
 P. Coloma, 
 JHEP 1603, 016 (2016),
arXiv:1511.06357.
\bibitem{nsi9}
 P. Coloma and T. Schwetz, 
 arXiv:1604.05772.
\bibitem{nsi10}
M. Masud and P. Mehta, 
arXiv:1606.05662.

\bibitem{dayabay}
 F. P. An et al. [Daya Bay Collaboration], 
 Phys. Rev. D 95, no. 7, 072006 (2017) [arXiv:1610.04802 [hep-ex]].


\bibitem{reno}
 S. H. Seo et al. [RENO Collaboration], 
Phys. Rev. D 98, no. 1, 012002 (2018) [arXiv:1610.04326
[hep-ex]].


\bibitem{Gando:2010aa}
A.~Gando \textit{et al.} [KamLAND],
Phys. Rev. D \textbf{83} (2011), 052002
[arXiv:1009.4771 [hep-ex]].





\bibitem{t2k}
K. Abe et al. [ T2K Collaboration ], Phys. Rev. Lett. 107,
041801 (2011). [arXiv:1106.2822 [hep-ex]].


\bibitem{Jun}
 H. T. J. Steiger [JUNO Collaboration], arXiv:1912.02038 [physics.ins-det].

\bibitem{degouvea}
A. de Gouvea and K. J. Kelly, Non-standard Neutrino Interactions at DUNE, Nucl. Phys. B 908 (2016) 318, arXiv:1511.05562



\bibitem{Giarnetti:2020bmf}
A.~Giarnetti and D.~Meloni,
[arXiv:2005.10272 [hep-ph]].

\bibitem{t2h}
KEK Preprint 2016-21 and ICRR-Report-701-2016-1, https://libextopc.kek.jp/preprints/PDF/2016/1627/1627021.pdf






\bibitem{liao}
J.~Liao, D.~Marfatia and K.~Whisnant,
JHEP \textbf{01} (2017), 071
[arXiv:1612.01443 [hep-ph]].








\bibitem{Adhikari:2012vc}
R.~Adhikari, S.~Chakraborty, A.~Dasgupta and S.~Roy,
Phys. Rev. D \textbf{86} (2012), 073010
[arXiv:1201.3047 [hep-ph]].






\bibitem{Girardi:2014kca}
I.~Girardi, D.~Meloni and S.~T.~Petcov,
Nucl. Phys. B \textbf{886} (2014), 31-42
[arXiv:1405.0416 [hep-ph]].






\bibitem{ponte}
B. Pontecorvo, 
Sov. Phys.
JETP 7 (1958) 172 173.

\bibitem{maki}
 Z. Maki, M. Nakagawa, and S. Sakata, 
 Prog. Theor. Phys. 28 (1962) 870 880.



\bibitem{Esteban:2019lfo}
I.~Esteban, M.~C.~Gonzalez-Garcia and M.~Maltoni,
JHEP \textbf{06} (2019), 055
[arXiv:1905.05203 [hep-ph]].






\bibitem{Coloma:2017ncl}
P.~Coloma, M.~C.~Gonzalez-Garcia, M.~Maltoni and T.~Schwetz,
Phys. Rev. D \textbf{96} (2017) no.11, 115007
[arXiv:1708.02899 [hep-ph]].
 

\bibitem{kopp}
J.~Kopp, M.~Lindner, T.~Ota and J.~Sato,
Phys. Rev. D \textbf{77} (2008), 013007
[arXiv:0708.0152 [hep-ph]].




\bibitem{Esteban:2020cvm}
I.~Esteban, M.~C.~Gonzalez-Garcia, M.~Maltoni, T.~Schwetz and A.~Zhou,
JHEP \textbf{09} (2020), 178
[arXiv:2007.14792 [hep-ph]].




\bibitem{Acciarri:2015uup}
R.~Acciarri \textit{et al.} [DUNE],
[arXiv:1512.06148 [physics.ins-det]].



\bibitem{Hyper-Kamiokande:2016dsw}
  [Hyper-Kamiokande Collaboration],
  KEK-PREPRINT-2016-21, ICRR-REPORT-701-2016-1.

\bibitem{Bakhti}
   P.~Bakhti and A.~Y.~Smirnov,
  arXiv:2001.08030 [hep-ph].



\bibitem{t2hk}
Y.~Itow \textit{et al.} [T2K],
[arXiv:hep-ex/0106019 [hep-ex]].




\bibitem{Bakhti:2013ora}
  P.~Bakhti and Y.~Farzan,
  JHEP {\bf 1310} (2013) 200
  [arXiv:1308.2823 [hep-ph]].

\bibitem{Bakhti:2014pva}
  P.~Bakhti and Y.~Farzan,
  JHEP {\bf 1407} (2014) 064
  [arXiv:1403.0744 [hep-ph]].







\bibitem{Huber:2004ka}
P.~Huber, M.~Lindner and W.~Winter,
Comput. Phys. Commun. \textbf{167} (2005), 195
[arXiv:hep-ph/0407333 [hep-ph]].

\bibitem{Huber:2007ji}
P.~Huber, J.~Kopp, M.~Lindner, M.~Rolinec and W.~Winter,
Comput. Phys. Commun. \textbf{177} (2007), 432-438
[arXiv:hep-ph/0701187 [hep-ph]].





















  \end{thebibliography}
\end{document}